\documentclass[acmsmall,screen,nonacm]{acmart}
\DeclareUnicodeCharacter{2260}{$\neq$}
\AtBeginDocument{%
  }

\newcommand{\ResIAConvs}{221}
\newcommand{\ResIAScrapeN}{34}
\newcommand{\ResIAScrapePct}{15.4}
\newcommand{\ResIACorrectionPct}{25.3}
\newcommand{\ResAlgHWN}{26}
\newcommand{\ResAlgHWPct}{30.6}
\newcommand{\ResCorpusHWPct}{17.6}
\newcommand{\ResHWUnionN}{79}
\newcommand{\ResHWAnnotatorN}{68}
\newcommand{\ResAlgPyIntroN}{44}
\newcommand{\ResAlgPyIntroPct}{51.8}
\newcommand{\ResAlgCorrectionPct}{25.9}
\newcommand{\ResCECorrectionPct}{35.7}
\newcommand{\ResCEConvs}{84}
\newcommand{\ResCEHWN}{18}
\newcommand{\ResCEHWPct}{21.4}
\newcommand{\ResNTOHWN}{17}
\newcommand{\ResNTOHWPct}{27.4}
\newcommand{\ResCTConvs}{6}
\newcommand{\ResCSConvs}{5}
\newcommand{\ResRecurCorpusConvs}{754}
\newcommand{\ResRecurCorpusIPs}{548}
\newcommand{\ResRecurLabeledIPs}{338}
\newcommand{\ResRecurMinIPs}{10}
\newcommand{\ResRecurChiSq}{14.91}
\newcommand{\ResRecurV}{0.21}
\newcommand{\ResRecurP}{.002}
\newcommand{\ResRecurIAK}{26}
\newcommand{\ResRecurIAN}{144}
\newcommand{\ResRecurIAPct}{18}
\newcommand{\ResRecurIAConvPct}{47}
\newcommand{\ResRecurCEK}{7}
\newcommand{\ResRecurCEN}{70}
\newcommand{\ResRecurCEPct}{10}
\newcommand{\ResRecurAlgK}{7}
\newcommand{\ResRecurAlgN}{77}
\newcommand{\ResRecurAlgPct}{9}
\newcommand{\ResRecurNTOConvs}{62}
\newcommand{\ResRecurNTOIPs}{62}
\newcommand{\ResHWChiSq}{24.82}
\newcommand{\ResHWP}{0.002}
\newcommand{\ResHWRows}{478}
\newcommand{\ResHWV}{0.23}

\newcommand{\MetPrelimConvs}{29}
\newcommand{\MetPrelimCodes}{11}
\newcommand{\MetPrelimFleiss}{0.52}
\newcommand{\MetPrelimPairMin}{0.44}
\newcommand{\MetPrelimPairMax}{0.65}
\newcommand{\MetCorpusConvs}{11,429}
\newcommand{\MetFenceConvs}{90,290}
\newcommand{\MetFencePython}{23,382}

\newcommand{\MetFenceJS}{7,327}
\newcommand{\MetScreenPairs}{2,080}
\newcommand{\MetScreenConvs}{1,712}
\newcommand{\MetConfirmedPairs}{1,039}
\newcommand{\MetConfirmedConvs}{957}
\newcommand{\MetAnchorConvs}{12}
\newcommand{\MetRoundTwoConvs}{30}
\newcommand{\MetRoundTwoCands}{35}
\newcommand{\MetHumanFleiss}{0.7205}
\newcommand{\MetHumanFleissLo}{0.56}
\newcommand{\MetHumanFleissHi}{0.83}
\newcommand{\MetHumanPairMin}{0.70}
\newcommand{\MetHumanPairMax}{0.75}
\newcommand{\MetOpusMissed}{4}
\newcommand{\MetOpusAdded}{5}
\newcommand{\MetSonnetMissed}{11}
\newcommand{\MetSonnetAdded}{1}
\newcommand{\MetOpusPairMin}{0.71}
\newcommand{\MetOpusPairMax}{0.76}
\newcommand{\MetHumanAgreeMin}{93.8}
\newcommand{\MetHumanAgreeMax}{95.4}
\newcommand{\MetOpusAgreeMin}{94.2}
\newcommand{\MetOpusAgreeMax}{95.4}
\newcommand{\MetReliabilityCells}{240}
\newcommand{\MetMajorityPos}{23}
\newcommand{\MetMajorityNeg}{217}
\newcommand{\MetFleissWithOpus}{0.7247}
\newcommand{\MetConfCalls}{2,080}
\newcommand{\MetConfTokIn}{11,540,331}
\newcommand{\MetConfTokOut}{270,630}
\newcommand{\MetAnnoCalls}{1,039}
\newcommand{\MetAnnoTokIn}{7,142,297}
\newcommand{\MetAnnoTokOut}{79,475}
\newcommand{\MetIARevUp}{105}
\newcommand{\MetIARevDown}{14}
\newcommand{\MetScreenTotalCalls}{91,432}
\newcommand{\MetAnnoPairsInitial}{548}
\newcommand{\MetAnnoConvsInitial}{511}
\newcommand{\MetAnnoPairs}{639}
\newcommand{\MetAnnoConvs}{600}
\newcommand{\MetAnnoErrCells}{4}
\newcommand{\MetVerifConvs}{953}
\newcommand{\MetAutoNonEnglish}{104}
\newcommand{\MetNonEnglishRestored}{1}
\newcommand{\MetManualNonEnglish}{23}
\newcommand{\MetNonEnglishExcluded}{126}
\newcommand{\MetDupCopies}{29}
\newcommand{\MetDupClusters}{16}
\newcommand{\MetNEDupOverlap}{3}
\newcommand{\MetFTClusters}{32}
\newcommand{\MetFTRemoved}{47}
\newcommand{\MetFTRemovedLabeled}{20}
\newcommand{\MetFTRemovedIA}{18}
\newcommand{\MetFTRemovedCE}{1}
\newcommand{\MetFTRemovedAlg}{1}
\newcommand{\MetCorpusFinal}{754}
\newcommand{\MetHumanOpusKappa}{0.87}
\newcommand{\MetVerifyKappaGL}{0.95}
\newcommand{\MetVerifyRelabelGL}{4}
\newcommand{\MetVerifyKappaSB}{0.97}
\newcommand{\MetVerifyRelabelSB}{2}
\newcommand{\MetVerifyKappaAPSA}{0.91}
\newcommand{\MetVerifyRelabelAPSA}{5}
\newcommand{\MetLabeledConvs}{450}
\newcommand{\MetCorpusUserTurns}{40,331}
\newcommand{\MetRoundTwoUserTurns}{134}

\newcommand{\nRefusals}{138}
\newcommand{\nLabeled}{450}
\newcommand{\pctRefusals}{30.7}
\newcommand{\refusalFamChiSq}{3.87}
\newcommand{\refusalFamP}{0.049}
\newcommand{\pctRefusalGptThree}{36.0}
\newcommand{\pctRefusalGptFour}{27.2}
\newcommand{\refusalLabelChiSq}{40.88}
\newcommand{\refusalLabelV}{0.29}
\newcommand{\refusalFamPhi}{0.09}
\newcommand{\refusalLabelP}{$1\times10^{-5}$}
\newcommand{\pctRefusalNontext}{62.9}
\newcommand{\pctRefusalInternet}{25.3}
\newcommand{\pctRefusalAlgebra}{25.9}
\newcommand{\nRefusalCT}{0}
\newcommand{\nRefusalCS}{0}

\newcommand{\SensNConv}{450}
\newcommand{\SensNIP}{338}
\newcommand{\SensNIPMulti}{45}
\newcommand{\SensNConvFromMulti}{157}
\newcommand{\SensFamGlmerP}{0.079}
\newcommand{\SensFamOR}{0.65}
\newcommand{\SensFamORLo}{0.40}
\newcommand{\SensFamORHi}{1.05}
\newcommand{\SensLabGlmerChiSq}{38.75}
\newcommand{\SensLabGlmerDf}{7}
\newcommand{\SensLabGlmerP}{$2\times10^{-6}$}
\newcommand{\SensSingleN}{425}
\newcommand{\SensSingleGlmerChiSq}{33.03}
\newcommand{\SensSingleGlmerP}{$3\times10^{-5}$}
\newcommand{\SensRecAdjChiSq}{15.88}
\newcommand{\SensRecAdjDf}{3}
\newcommand{\SensRecAdjP}{0.001}

\newcommand{\wordnum}[1]{\ifcase#1 zero\or one\or two\or three\or four\or five\or six\or seven\or eight\or nine\else #1\fi}
\makeatletter
\def\sb@pz#1#2\@nil{\ifx.#1 0.#2\else#1#2\fi}
\newcommand{\pz}[1]{\begingroup\edef\sb@tmp{#1}\expandafter\sb@pz\sb@tmp\@nil\endgroup}
\makeatother

\usepackage{tikz}
\usetikzlibrary{positioning,arrows.meta}

\begin{document}

\title{User Misconceptions of LLM-Based Conversational~Programming~Assistants}
\titlenote{An earlier version of this work, presenting a preliminary codebook, appeared at the ICSE 2026 Journal Ahead Workshop (JAWs) \cite{OBrien2026UserAssistants}.}

\author{Gabrielle O'Brien}
\orcid{0009-0001-3198-3586}
\affiliation{%
  \institution{University of Michigan}
  \city{Ann Arbor}
  \state{Michigan}
  \country{USA}
}
\email{elleobri@umich.edu}

\author{Antonio Pedro Santos Alves}
\orcid{0000-0002-3415-6201}
\affiliation{%
  \institution{Pontifical Catholic University of Rio de Janeiro}
  \city{Rio de Janeiro}
  \country{Brazil}}
\email{apsalves@inf.puc-rio.br}

\author{Sebastian Baltes}
\orcid{0000-0002-2442-7522}
\affiliation{%
  \institution{Heidelberg University}
  \city{Heidelberg}
  \country{Germany}}
\email{sebastian.baltes@uni-heidelberg.de}

\author{Grischa Liebel}
\orcid{0000-0002-3884-815X}
\affiliation{%
  \institution{Reykjavik University}
  \city{Reykjavik}
  \country{Iceland}}
\email{grischal@ru.is}

\author{Marcos Kalinowski}
\orcid{0000-0003-1445-3425}
\affiliation{%
  \institution{Pontifical Catholic University of Rio de Janeiro}
  \city{Rio de Janeiro}
  \country{Brazil}}
\email{kalinowski@inf.puc-rio.br}

\begin{abstract}
Programming assistants powered by large language models (LLMs) have become widely available, with conversational assistants such as ChatGPT particularly accessible to novice programmers. However, varied tool capabilities and inconsistent availability of extensions (e.g., web search, code execution, retrieval-augmented generation) create opportunities for user misconceptions that may lead to over-reliance, unproductive practices, or insufficient quality control. 
We characterize the misconceptions that users of conversational LLM-based assistants may hold in programming contexts. We screened \MetCorpusConvs{} Python-related conversations from the openly available WildChat dataset with a validated LLM annotation pipeline, then hand-annotated the \MetCorpusFinal{} candidate conversations it flagged. Of these, \MetLabeledConvs{} contain a prompt consistent with at least one of eight potential misconceptions: misplaced expectations about capabilities such as web access, code execution, non-text outputs, and session memory. We also characterize how the assistant responds when a prompt presupposes a capability it lacks: responses range from explicit refusal through qualified answers to fabricated compliance, and explicit refusals appear in only a minority of labeled conversations. Among the most frequent misconceptions, explicit refusals are rarest where compliance is easiest to fabricate. Our findings reinforce the need for LLM-based tools to communicate their capabilities to users through channels other than the conversation itself.
\end{abstract}



\maketitle

\section{Introduction}

Programming assistants powered by large language models (LLMs) have rapidly proliferated across software development practices. Several studies indicate their potential to improve developer productivity \cite{Ziegler2022ProductivityCompletion, Weber2024SignificantModels}, though this finding is far from universal \cite{Stray2025DeveloperStudy, BeckerMeasuringProductivity}. Users employ these tools for diverse programming tasks, including information retrieval, code generation, debugging, documentation writing, and learning new technologies \cite{Liang2024AChallenges}. The range of available tools is broad and growing: interfaces range from auto-complete engines embedded in development environments to chat interfaces that mimic natural language conversations \cite{Treude2025HowEngineering}, with the latter proving particularly accessible to programmers across skill levels \cite{StackOverflow2025Survey, Fawzy2025VibeReview}.

A key risk for all users, especially novice programmers, is over-reliance. \citeauthor{Chen2021EvaluatingCode} name it as a key risk of code generation models, since generated code can appear correct without performing the intended task \cite{Chen2021EvaluatingCode}. In lab studies of Copilot, participants accepted generated code without validating it \cite{Vaithilingam2022ExpectationModels}, and first-time users were more over-reliant than experienced ones \cite{Barke2023GroundedModels}. For example, some programmers rely on asking the model to explain its code choices as their entire validation procedure \cite{Fawzy2025VibeReview, OBrien2025HowProgram}, even though LLM-generated explanations may be incorrect \cite{Kabir2024IsQuestions, Sarsa2022AutomaticModels}---even about code the LLM has just produced. In some contexts, users may not validate outputs at all. Novices accepted Copilot suggestions that they did not understand \cite{Prather2024TheProgrammers} and expected it to always generate correct code \cite{Prather2023ItsProgrammers}. Students using a classroom assistant noted that it ``can lie to you, and still sound confident'' \cite{Kazemitabaar2024CodeAid:Needs}, and vibe coders reviewed generated code minimally or trusted it uncritically \cite{Fawzy2025VibeReview}. Higher confidence in the model's abilities was associated with less critical thinking \cite{Lee2025TheWorkers}.

How programmers perceive the abilities and mechanisms behind LLM-based programming tools likely determines their confidence in these tools. The same holds beyond programming. People advised by an AI agent decide better when they perceive its strengths and weaknesses accurately \cite{Kelly2023CapturingApproach}, and team performance depends on how well the human understands the AI's error boundary \cite{Bansal2019BeyondPerformance}. Users of an agentic chat assistant anchored their confidence on its search-like behavior while lacking insight into the generative model behind it \cite{Brachman2025BuildingChat-bot}. This includes forming mental models of tool capabilities and limitations \cite{Prather2024TheProgrammers}. For example, a user may trust generated answers about a programming library more if they believe a tool can search official documentation rather than relying on a frozen training dataset. Similarly, if a user expects a tool to execute tests, they may be more confident in accepting generated programs. However, these capabilities can differ widely across tools with similar interfaces. In the present study, we investigate what users of conversational LLM-based assistants (such as ChatGPT \cite{Treude2025HowEngineering}) may misunderstand about how these tools work. Because a user's mental model is shaped not only by their expectations but also by the feedback the tool provides, we also examine the assistant's side of these exchanges. Specifically, we ask:

\begin{description}
    \item[RQ1] What potential misconceptions about a conversational assistant's capabilities are visible in real programming-related chat logs, and in what contexts do they arise?
    \item[RQ2] When a prompt presupposes a capability the assistant lacks, how does the assistant respond---and how clearly do its responses communicate the actual capability boundary?
\end{description}

To address RQ1, we conduct a qualitative analysis of the openly available WildChat dataset \cite{Zhao2024WildChat:Wild}, building a codebook of eight kinds of potential user misconceptions in the context of programming-related conversations with a conversational assistant. Using an LLM screening stage followed by model-assisted, human-verified annotation, we processed \MetCorpusConvs{} Python programming-related conversations from the dataset. We find evidence of misconceptions about the availability of features such as web access and code execution, knowledge cutoffs, the models' ability to produce non-text outputs, and access to the users' local machine. For RQ2, we characterize the assistant's responses to these requests, which range from explicit refusal through answering with qualifications to fabricated compliance---patterns that differ in how much they reveal about what the system can actually do. We conclude by discussing implications for designing LLM-based tools that better communicate their capabilities to users. This article extends a workshop paper \cite{OBrien2026UserAssistants} that cataloged candidate misconceptions and developed an initial codebook. New here are the refinement of that codebook to achieve higher inter-rater reliability, the two-stage LLM screening and annotation pipeline with human verification of every candidate conversation (\MetCorpusFinal{} conversations, yielding the \MetLabeledConvs{} labeled conversations we report), the qualitative account of each misconception, the analysis of assistant responses (RQ2), an analysis of repeated attempts by the same user, and sensitivity analyses of the statistical results.

\section{Related Work}
There is a well-documented tendency to over-rely on automated systems (``automation bias''). A systematic review found the effect robust across research fields \cite{Goddard2012AutomationMitigators}, and it has been observed in a range of settings beyond the laboratory \cite{Gadala2017AutomationStrategies}. Beginning programmers exhibited it when they assumed LLM-generated code was correct even after being told it might contain bugs \cite{Zi2025ICode}. Even before the LLM era, studies of end-user spreadsheet programmers found that people were far more likely to judge an incorrect computed value correct than a correct value incorrect, and that simply seeing a spreadsheet produce values, with no indication of whether they were right, made users more confident in it \cite{KoTheEngineering}.

Similarly, user studies with program synthesizers indicate a propensity for overconfidence: \citeauthor{Ferdowsifard2020Small-stepExample} introduced the term ``user-synthesizer gap'' for the gap between a user's mental model of a synthesizer's abilities and its actual abilities \cite{Ferdowsifard2020Small-stepExample}, and \citeauthor{Jayagopal2022ExploringProgrammers} observed a ``monotonicity belief'' that giving a synthesizer more information always improves its results \cite{Jayagopal2022ExploringProgrammers}. Indeed, programmers who understand what tasks a synthesizer can accomplish know when to invoke it rather than try alternative approaches \cite{Ferdowsifard2020Small-stepExample}. However, those with poor mental models---particularly those overestimating capabilities---may waste time trying to accomplish impossible tasks. Correspondingly, studies of AI-aided decision making have found that users with more accurate mental models achieve better outcomes. Players who estimated an AI agent's abilities more accurately won more often \cite{Gero2020MentalSetting}, and users who most improved their mental model of a recommender were more likely to make it operate to their satisfaction \cite{Kulesza2012TellAgent}. A mismatch between a person's mental model and an AI's actual error boundary leads to suboptimal decisions \cite{Bansal2019BeyondPerformance}.

Previous user-interaction literature suggests that neural-network-based tools pose exceptionally high barriers for developing working mental models. Even for a simple cooperative game agent, players' estimates of the agent's abilities varied, and better estimates went with better outcomes \cite{Gero2020MentalSetting}. Non-experts expected plain descriptions of desired behavior to be ``understood'' and acted upon, and read failures as limits of the model rather than of the prompt \cite{Zamfirescu-Pereira2023WhyPrompts}. Output failures can stem from the prompt, the system's non-determinism, or its training data without an obvious way to tell which \cite{Tankelevitch2024TheAI}. These challenges are compounded by LLMs' tendency to misrepresent their own processes---as one recent technical report noted, ``we found that o3 frequently fabricates actions it took to fulfill user requests, and elaborately justifies the fabrications when confronted by the user'' \cite{Chowdhury2025InvestigatingAI}. 

Consistent with this, when exploring mental models of an agentic LLM-based assistant for information-seeking in non-programming contexts, users commonly expected rigid, rule-based behavior (e.g., decision trees or database lookups) rather than probabilistic inference for tool calling \cite{Brachman2025BuildingChat-bot}. Users also expected responses to reflect information retrieval actions rather than training data. Therefore, understanding information provenance was closely related to trust: Users wanted to know sources before trusting responses in high-stakes situations \cite{Brachman2025BuildingChat-bot}.

The computing education literature on programmer misconceptions, though it predates LLM-based tools, offers a vocabulary for describing them. \citet{Sorva2013NotionalEducation} defined misconceptions as ``understandings that are deficient or inadequate for many practical programming contexts''. Programmer misconceptions about computers, languages, and complex systems have been studied both as targets for educational intervention and to understand barriers to student achievement. \citeauthor{Iii1994MisconceptionsTransition} questioned the view of misconceptions as flawed ideas that instruction must confront and replace \cite{Iii1994MisconceptionsTransition}. \citeauthor{Qian2017StudentsProgramming} reviewed the misconceptions of introductory programming students \cite{Qian2017StudentsProgramming}, and \citeauthor{Oliveira2023StudentMisconceptions} studied students' misconceptions about code refactoring \cite{Oliveira2023StudentMisconceptions}. 

Prior analyses of developer--assistant chat logs mostly catalog what developers ask for. DevGPT collected 29,778 prompt--response pairs from ChatGPT links shared on GitHub and Hacker News \cite{Xiao2024DevGPTConversations}; follow-up studies classified its conversations into task types and compared conversation length across them \cite{Champa2024ChatGPTDevelopment}, or identified 19 conversation topics, one of them a ``Turing test'' category in which developers probed the model's reasoning \cite{Mohamed2024ChattingChatGPT}. \citet{Tang2026ProgrammingSessions} labeled 74,998 developer messages to IDE-integrated assistants by behavioral intent. Outside programming, a log analysis of Google Bard interactions noted in passing that users asked the system about its capabilities and that responses sometimes described abilities the system did not have \cite{Trippas2024WhatWild}. Labeling logs at this scale often involves automated classification, which has its own methodological literature on maintaining rigor \cite{Dunivin2025ScalingAnalysis, Shankar2024WhoPreferences, Ernst2026GenAIEngineering}.

Two further strands motivate our questions. Assistants tend toward sycophancy \cite{Sharma2023TowardsModels}, and reinforcement learning from human feedback makes responses more convincing without making them more correct \cite{Wen2025LANGUAGERLHF}, so an assistant may accommodate a mistaken assumption rather than correct it. On the user side, forming a mental model of what an LLM-based tool can do is itself difficult \cite{Sarkar2022WhatIsIt, Liu2023WhatItWants, Tankelevitch2024TheAI}; design guidance has long asked systems to make this clear \cite{Amershi2019GuidelinesInteraction}, and interviewed developers described learning tool capabilities partly from other users' shared experiences \cite{Cheng2024ItTools}. To our knowledge, no prior work has measured capability presuppositions in real programmer--assistant logs at scale, nor characterized how the assistant responds when a prompt presupposes a capability it lacks.

\section{Research Approach}\label{sec:research-approach}
We study the misconceptions that programmers may have about LLM-based tools. We begin by articulating our assumptions and approach.

\subsection{Defining Misconceptions}
We define user ``misconceptions'' as instances where a user erroneously believes a tool has an affordance it lacks (or fails to perceive a critical property). In tool design, affordances are ``those fundamental properties that determine just how the thing could possibly be used'' \cite{Norman1988TheThings}. Affordances may be real or erroneously perceived if a user's mental model is incomplete or faulty \cite{Norman1999AffordanceDesign}.
For example, a user prompting a conversational assistant might perceive the tool has a web plugin that retrieves information from official documentation before generating responses. If the tool lacked web search capabilities, we would consider this a misconception. This would be relevant to the programmer's confidence in the LLM, as they may weight certainty differently if they believe the response summarizes recent official documentation rather than relying solely on frozen training data. Throughout, we use \emph{assistant} (short for conversational assistant) for the deployed system a user interacts with, including its interface and configuration, and \emph{model} for the underlying LLM that generates its responses. We use \emph{code} for each of the eight misconception types in our codebook and \emph{label} for the assignment of a code to a conversation.

Misconceptions are context-specific because conversational assistant affordances vary across providers and versions. For example, OpenAI released a Python interpreter for GPT-4 in 2023, initially as an opt-in feature for Plus subscribers, later transitioning to a default feature \cite{OpenAI2023CodeInterpreter}. Thus, expecting a code interpreter may or may not be a misconception depending on which tool is used, when, and at what subscription tier. A misconception could also reflect insensitivity to a tool property. For example, a user might not perceive that a given LLM-based tool is non-deterministic, becoming confused when the same prompt yields different responses.

\subsection{Our Approach}
When we first set out to undertake this study, our main question was, \textit{what kinds of programming-related user misconceptions are visible in chat logs?} However, we realized there was also a substantial measurement challenge, which we detailed in the Journal Ahead Workshop (JAWs) version of this study \cite{OBrien2026UserAssistants}. The present paper is a continuation of that work, addressing the measurement issues we faced. For space reasons we do not reproduce the preliminary study here. Appendix~\ref{sec:preliminary-study} documents the brainstormed candidate themes (Table~\ref{tab:brainstormed-themes}), the first version of the codebook (Table~\ref{tab:preliminary-codebook}), and the preliminary coding procedure.

In our initial approach, we brainstormed the misconceptions a user might have when interacting with conversational LLMs based on our knowledge of the literature and our own interactions with students and fellow programmers. Then, we looked at a sample of 500 randomly selected Python programming conversations from the WildChat dataset. After several rounds of open coding and axial coding, we conducted a validation round on a stratified subset of \MetPrelimConvs{} conversations with three additional coders (all authors on the original work)---and found only moderate inter-rater reliability \cite{Landis1977TheData}, even after discussion. When we compared the four resulting label sets---the three validation coders' labels and the coding team's adjudicated labels---pooled over the conversation-by-code grid (\MetPrelimConvs{} conversations $\times$ \MetPrelimCodes{} codes; Table~\ref{tab:preliminary-codebook}), Fleiss' $\kappa$ was \MetPrelimFleiss{}, with pairwise Cohen's $\kappa$ between \MetPrelimPairMin{} and \MetPrelimPairMax{}.

This exercise revealed several issues. First, some of the misconception codes we drafted were not readily visible in logs. We could not see discernible evidence for users being surprised by non-determinism (a brainstormed theme, Table~\ref{tab:brainstormed-themes}, that never entered the codebook), for example, because we didn't see many conversations where users would repeat the same prompts. Additionally, some kinds of misconceptions in our codebook were simply too hard to map to users' own words with much certainty. For example, one of our codes was for ``dynamic analysis'', in which the user appears to have a misplaced expectation that the LLM can review whether a program ``works''---which would typically require more than a static analysis of the program. Raters disagreed about the use of highly colloquial language---we did not feel confident as a group that asking a chatbot if a program ``works'' implied any particular mental model.

Other issues related to unresolved disagreements about how to handle cases like unclear language (some conversations switched languages partway through; the WildChat dataset has machine-applied language tags that are imperfect) or prompts that might be homework assignments. Some disagreements also occurred simply because conversations can be very long, involving many user turns, so spotting a misconception present in a single sentence can require reading the entire transcript. Raters might disagree simply because of accidentally overlooking a phrase in a long transcript.

To continue the work in a direction where we could be more confident what we were measuring, we made several design choices: first, we limited ourselves to investigating \textit{only} eight kinds of misconceptions that our initial investigation turned up clear, unambiguous cases of, where the six preliminary-study annotators broadly agreed on the definition and at least one anchor case could be found. These were codes like ``internet access'', where a user would include a URL in their prompt with instructions to access data from there, or ``non-text output'', where users requested diagrams or plots (seemingly outside of a context where the LLM typically responded with code snippets)---features that are more straightforwardly observed from a conversational log. Second, we implemented an LLM-based screening procedure to permissively flag candidate conversations from the larger dataset, helping us focus our investigation on a sample that is more likely to be informative (as we learned in the preliminary work, where only 50 of the 500 sampled conversations were flagged as containing evidence for one of the misconceptions). Finally, we used a human-LLM hybrid coding approach to annotate the candidates according to our codebook. We then present key themes in how the potential misconceptions surface in logs. 

\subsection{Methodological Limitations}
Our approach is indirect because we cannot directly measure whether users understand a conversational assistant's affordances. It is also impossible to be certain what misconceptions users have from conversational logs alone, as users might experiment with impossible prompts while exploring the tool. Therefore, what we code is the \emph{presupposition of an absent affordance} in a prompt---our observable proxy for a potential misconception (e.g., providing a URL to a model without browser access). For brevity, we sometimes use ``misconception'' in our results, but this should be understood as inference from that proxy rather than direct measurement. Acknowledging this limitation, we believe there is value in cataloging potential misconceptions users may have and the programming contexts where they arise.

Additionally, by focusing on interactions with a generic conversational assistant \cite{Treude2025HowEngineering} rather than developer tools, we expect conversations may skew towards what would now be called \textit{vibe coding} interactions. ``Vibe coding'', a term coined by \citeauthor{Karpathy2025TheresCoding.} for programming in which one ``fully give[s] in to the vibes'' and ``forget[s] that the code even exists'' \cite{Karpathy2025TheresCoding.}, describes users relying heavily on natural language to specify programming requirements while reviewing the generated code minimally \cite{Fawzy2025VibeReview}, prioritizing flow and experimentation over precision or control \cite{Pimenova2025GoodCoding}. Although conversational assistants are often viewed as especially accessible to beginners \cite{Fawzy2025VibeReview, Treude2025HowEngineering}, WildChat is anonymous, so we cannot confirm users' professions or backgrounds.

\section{Methods}\label{sec:methods}
Figure~\ref{fig:pipeline} summarizes the data pipeline described in this section, from corpus construction through label verification and cleaning.

\begin{figure}[t]
  \centering
  \begin{tikzpicture}[
      node distance=5.5mm,
      stage/.style={draw, semithick, align=center, text width=98mm,
                    inner sep=5pt, font=\small},
      llmstage/.style={stage, fill=black!8},
      flow/.style={-{Stealth[length=2.2mm]}, semithick},
      steplabel/.style={font=\footnotesize\itshape, align=center, midway, right=1mm}
    ]
    \node[stage] (corpus) {\textbf{Corpus construction}\\
      WildChat-1M (1,039,785 conversations) filtered to English conversations
      with at least one Python code block: \MetCorpusConvs{} conversations};
    \node[llmstage, below=of corpus] (flagging) {\textbf{Candidate flagging}\\
      A deliberately permissive open-weight screen (\texttt{gpt-oss-120b})
      flags \MetScreenPairs{} candidate conversation--code pairs in
      \MetScreenConvs{} conversations for the 8 codes;
      \texttt{claude-sonnet-4-6} re-scores them, keeping \MetConfirmedPairs{}
      pairs in \MetConfirmedConvs{} conversations};
    \node[stage, below=of flagging] (irr) {\textbf{Human annotation and codebook refinement}\\
      3 annotators label a stratified subset of \MetRoundTwoConvs{}
      conversations (\MetRoundTwoCands{} candidate codes); disagreements
      reviewed and exclusion rules added; pooled Fleiss
      $\kappa = \MetHumanFleiss$ (0.62 before discussion)};
    \node[llmstage, below=of irr] (opus) {\textbf{Model-assisted annotation}\\
      \texttt{claude-opus-4-7}, prompted with the finalized codebook,
      adjudicates all \MetConfirmedPairs{} candidate pairs: \MetAnnoPairs{}
      labels in \MetAnnoConvs{} conversations};
    \node[stage, below=of opus] (verify) {\textbf{Label verification}\\
      One author reviews every candidate conversation, accepting or rejecting
      each model-applied code and adding missed codes
      ($\kappa = \MetHumanOpusKappa$ with the model); three co-authors
      re-review segments of 100 conversations each};
    \node[stage, below=of verify] (clean) {\textbf{Cleaning}\\
      Language exclusions and de-duplication reduce the \MetVerifConvs{}
      reviewed conversations to \MetCorpusFinal{}, of which
      \MetLabeledConvs{} carry at least one code};
    \draw[flow] (corpus) -- (flagging);
    \draw[flow] (flagging) -- (irr);
    \draw[flow] (irr) -- (opus);
    \draw[flow] (opus) -- (verify);
    \draw[flow] (verify) -- (clean);
  \end{tikzpicture}
  \Description{Flowchart of six stages connected top to bottom by arrows: corpus construction from WildChat-1M; candidate flagging by an LLM screen and re-scoring; human annotation and codebook refinement; model-assisted annotation with the finalized codebook; label verification by the authors; and cleaning, which yields the analysis corpus. The two LLM stages are shaded.}
  \caption{Data pipeline. Shaded stages were performed by an LLM; unshaded
  stages by the authors. After cleaning (Section~\ref{sec:corpus-cleaning}),
  the analysis corpus contains \MetCorpusFinal{} conversations,
  \MetLabeledConvs{} with at least one misconception code.}
  \label{fig:pipeline}
\end{figure}

\subsection{Data}
A challenge for studying how users interact with conversational LLM-based tools in realistic settings is that many such datasets are proprietary. We selected the openly available WildChat corpus of over 1 million conversations between anonymous users and a free, publicly available conversational assistant~\cite{Zhao2024WildChat:Wild}. We also considered LMSYS-Chat-1M \cite{Zheng2024LMSYSChat1M}, but this dataset emphasized a ``gamified'' interface inviting head-to-head comparisons of 25 LLMs. We selected WildChat because its user interface seemed more ecologically valid for our purposes.
As reported by \citet{Zhao2024WildChat:Wild}, WildChat data was collected between April 2023 and May 2024, with 2,713,695 turns in 1,039,785 conversations. Users interacted with a chat interface hosted on Hugging Face\footnote{\url{https://huggingface.co/spaces}} supported by OpenAI's GPT-3.5-Turbo and GPT-4 APIs. As a mandatory condition to access the assistant, users agreed to share chat transcripts, IP addresses, and request headers. Users who affirmatively consented to data sharing could access the service free of charge.

Critically, the WildChat assistant's API configuration is public: the data collection is described in \citet{Zhao2024WildChat:Wild}, and the source of the Hugging Face Space that served it \cite{Deng2023ChatGPTSpace} passes only the model name, message history, and sampling parameters to the chat completions endpoint, with no tools, file inputs, or browsing enabled. The API calls did not allow the assistant to support web search, file uploads, non-text inputs/outputs, or ``tool calling'' (such as code interpreters). Each conversation was a distinct session; users could not create accounts, so the chatbot had no ``memory'' of previous conversations. Additionally, the deployment accepted and returned text only.
Based on IP addresses, users were located predominantly in the US (21.6\%), Russia (15.6\%), and China (10.0\%). \citeauthor{Zhao2024WildChat:Wild}'s analysis of 1,000 random prompts indicated 6.7\% were coding-related (other categories: assisting/creative writing 61.9\%, analysis/decision explanation 13.6\%, factual info 6.3\%). English was the most common prompt language (53\%), followed by Chinese (13\%) and Russian (12\%).

\subsubsection{Filtering}
To focus on programming contexts, we identified coding-related conversations by
applying a regular expression filter for code block formatting characters in
user prompts and chatbot responses, which for GPT models typically carry a
language tag. Out of \MetFenceConvs{} conversations with code block formatting,
Python was the most common language tag (\MetFencePython{} conversations),
followed by JavaScript (\MetFenceJS{}); no other language tag appeared in more
than 5,000 conversations. We restricted the analysis to Python, which was both the
most common and the language all authors knew. We rebuilt the corpus from the raw WildChat-1M
release with a scripted, reproducible filter---conversations machine-tagged as
English containing at least one Python code block in any message---yielding
\MetCorpusConvs{} conversations (\MetCorpusUserTurns{} user turns). The English tag is WildChat's conversation-level language field, and the Python test matches a tagged code fence in either a user or an assistant message; conversations whose Python code was pasted without a language tag are therefore not in the corpus.

\subsection{Initial Codebook Development}
We developed our initial codebook in the work presented at JAWs \cite{OBrien2026UserAssistants}; the full method is repeated in Appendix~\ref{sec:preliminary-coding}, and the preliminary codebook is reproduced in Table~\ref{tab:preliminary-codebook}.
From this codebook, we rescoped our analysis only to codes where we found reasonably high agreement from the six annotators in the preliminary stage that a) the definition of the code was reasonable, and b) at least one anchoring incident could be found as a reference for this occurring. After rescoping, we updated the codebook with definitions, anchor cases, and boundary conditions (what to exclude) and reviewed as a group asynchronously.

\subsection{Identifying Candidate Conversations} 
Conversations that suggest misconceptions are rare in the dataset, and the preliminary study involved several annotators looking through conversational logs in which the negative class dominates. To narrow down the search space, we wrote a deliberately permissive prompt for an LLM pipeline to flag candidate conversations for each code. In other words, we instructed the LLM to flag even likely ambiguous matches to the code criteria. Following guidelines for improving the performance of LLMs on qualitative analysis tasks, we required the model to produce its reasoning \textit{before} stating its verdict \cite{Dunivin2025ScalingAnalysis}.   Our goal here was to collect a reasonably focused sample of data for qualitative analysis. We accept that through this approach we cannot precisely know how many ``true positives'' in the full corpus were screened out. Appendix~\ref{sec:pipeline-details} reports the operational detail of the LLM stages---prompt structure, models and endpoints, sampling parameters, run stability, and inference dates---following the reporting guidelines of \citet{Baltes2025GuidelinesModels}; the complete prompts are included in the replication package.
\subsubsection{Validating the Filter Prompts} We assembled an anchor set of \MetAnchorConvs{} conversations---one or two consensus examples for each of the eight misconception codes---and iterated on filter prompt language until the pipeline reliably recovered all \MetAnchorConvs{}. Because the same anchors guided the prompt revisions, this is recall on the tuning set rather than on held-out data. The prompts pair each code's definition with quoted example phrases drawn from the anchors; to rule out string matching, we confirmed that replacing those phrases with paraphrases did not affect recall of the \MetAnchorConvs{} target conversations.

\subsubsection{Model Selection and Screening} We conducted the anchor validation with \verb|claude-opus-4-7| via a university-hosted endpoint, but running it across the corpus (\MetCorpusConvs{} $\times$ 8 prompts) was projected to cost roughly \$2{,}000, which was prohibitively expensive. We evaluated less expensive models on the anchor set (the candidates are listed in Appendix~\ref{sec:pipeline-details}) and found that \verb|gpt-oss-120b|, an open-weight model hosted on the university HPC cluster, recovered all \MetAnchorConvs{} anchors in each of ten repeated runs, though the set of other cells it flagged varied from run to run (Appendix~\ref{sec:pipeline-details}). The corpus was screened once, so a repeat run might yield a somewhat different candidate set. We also noted that in spot checks of the model's reasoning traces, \verb|claude-sonnet-4-6| (another university-hosted endpoint) had notably higher-quality reasoning. For example, it could discern that a user asking about train/test splits was not asking for an algebra problem to be solved, but asking about a convention in machine learning. To keep costs moderate, we therefore ran two rounds of screening: first, we ran the candidate-flagging pipeline with \verb|gpt-oss-120b|, and then ran any candidate pairs (\MetScreenPairs{} candidate code labels across \MetScreenConvs{} conversations) it surfaced through \verb|claude-sonnet-4-6| for another label. Where disagreements occurred, we used the label from \verb|claude-sonnet-4-6|. This left us with a pool of \MetConfirmedPairs{} candidate pairs across \MetConfirmedConvs{} conversations.

\subsection{Human Validation and Codebook Finalization}
The goal of the flagging pipeline was to narrow the set of conversations that humans would read, rather than to apply codes directly. At this point, we conducted another round of human annotation on a new subset of \MetRoundTwoConvs{} conversations (with \MetRoundTwoCands{} misconception code candidates) from the flagged set. These conversations contained \MetRoundTwoUserTurns{} user turns. The subset was chosen with stratification, guaranteeing that at least two candidates for every code were present to ensure coverage.

Using a browser-based annotation tool we created, three annotators (GO, SB, and GL; all annotators in this study are authors and software engineering researchers with Python experience) decided what labels should be applied to the conversations, which could include adding any labels that the pipeline had not applied (though this happened in only one case). Initially, before any discussion, pooled Fleiss' $\kappa$ was 0.62. In reviewing one another's comments, we quickly resolved two disagreements that stemmed from missed language in the conversation (annotator lapses of attention). There were also disagreements because one rater had coded requests for equations as ``Algebra'', whereas the others considered asking for equations (``What is the equation for....'') separate from asking for equations to be solved. We therefore ruled that the codebook should be updated such that asking for an equation on its own is not part of the ``Algebra'' code, and added this to the code's exclusion list. After these changes (post-discussion, under the revised codebook), we had a pooled $\kappa = \MetHumanFleiss$ (95\% bootstrap CI \MetHumanFleissLo{}--\MetHumanFleissHi{}, resampling conversations), with pairwise $\kappa$ between \MetHumanPairMin{} and \MetHumanPairMax. By the conventional benchmarks \cite{Landis1977TheData}, $\kappa$ in this range indicates substantial agreement, which we treated as sufficient to proceed. The full boundary conditions for every code are reproduced in Appendix~\ref{sec:code-boundaries}. 

\subsection{Model-Assisted Annotation}
With the codebook finalized, we built a pipeline to have an LLM-annotator adjudicate every candidate code flagged in the screening stage. Using prompt templates built from our finalized codebook and any new boundary/exclusion cases, we compared how \verb|claude-sonnet-4-6|  and \verb|claude-opus-4-7| would assign codes to the \MetRoundTwoConvs{} conversations just reviewed in the human-annotation stage. Opus' performance was well within the human-human agreement band on both raw agreement and chance-corrected agreement (raw agreement with each rater \MetOpusAgreeMin{}--\MetOpusAgreeMax\%, versus \MetHumanAgreeMin{}--\MetHumanAgreeMax\% between human pairs; pairwise $\MetOpusPairMin \leq \kappa \leq \MetOpusPairMax$) and adding it as a fourth rater raised the pooled Fleiss' $\kappa$ from \MetHumanFleiss{} to \MetFleissWithOpus{}. It was therefore at least as consistent with the group as every individual member. Sonnet had similar overall accuracy; however, it was far more conservative applying labels than the human raters were---it missed \MetSonnetMissed{} of the \MetMajorityPos{} codes applied by majority vote across the \MetReliabilityCells{} cells, and added \MetSonnetAdded{} among the \MetMajorityNeg{} cells where the majority applied none. Sonnet also seemed to miss some conversations where there was unanimous or near-unanimous agreement by the human raters. Opus, on the other hand, had a fairly balanced pattern of disagreements (\wordnum{\MetOpusMissed} of \MetMajorityPos{} missed, \wordnum{\MetOpusAdded} of \MetMajorityNeg{} added). Even though it was more expensive to run the annotation round with Opus, we chose it.

We then ran this pipeline over all \MetConfirmedPairs{} candidate pairs. The initial run applied \MetAnnoPairsInitial{} labels across \MetAnnoConvsInitial{} conversations. Early in the verification stage described below, we observed that the model had interpreted the Internet Access prompt overly literally: When a user supplied a link (for example, to a documentation page) without expressly instructing the model to go there, it did not apply the code, although such cases were within the code's intended scope. We revised the prompt, tested it on the conversations already labeled to confirm the revision fixed the problem, and re-ran the Internet Access candidate pairs before continuing verification; the re-run added the code to \MetIARevUp{} pairs and removed it from \MetIARevDown{}. The final model-applied label set contained \MetAnnoPairs{} labels across \MetAnnoConvs{} conversations. A further \wordnum{\MetAnnoErrCells} candidate pairs (each the only candidate in its conversation) could not be scored: Across repeated attempts, two transcripts were rejected by a web application firewall on the university API gateway and two received responses suppressed by the provider's content filter. These \wordnum{\MetAnnoErrCells} conversations were dropped from the analysis.

\subsection{Label Verification}\label{sec:corpus-cleaning}
At least one human annotator then reviewed every candidate conversation. Using the browser-based tool (Appendix~\ref{sec:verification-tool}, Figure~\ref{fig:annotation-tool}), one author (GO) reviewed each conversation alongside the model's per-code verdicts and reasoning, accepting or rejecting each candidate code and applying any codes the pipeline had missed. Agreement between the final human label set and the model-applied labels was $\kappa = \MetHumanOpusKappa$. Three co-authors (GL, SB, and APSA) each independently re-reviewed a unique segment of 100 conversations from the same review queue, viewing each conversation alongside GO's and the model's labels and flagging any cases where they disagreed with how the codes had been applied. Pooled $\kappa$ agreement between each reviewer and GO's labels was $\kappa = \MetVerifyKappaGL$ (GL), $\kappa = \MetVerifyKappaSB$ (SB), and $\kappa = \MetVerifyKappaAPSA$ (APSA). These are agreement rates against labels the reviewers could see, not independent codings, and should be read as a check on the verified labels rather than as inter-rater reliability. The three segments were consecutive blocks of the first 300 positions in the randomly ordered verification queue. GO then revisited every case where a disagreement occurred and discussed with the other annotators to resolve the label. This process triggered \wordnum{\MetVerifyRelabelGL} relabelings in GL's segment, \wordnum{\MetVerifyRelabelSB} in SB's, and \wordnum{\MetVerifyRelabelAPSA} in APSA's. Based on discussion, we further clarified rules for handling edge cases. For example, we did not consider prompts that appeared to be a formulaic jailbreak attempt as misconceptions, and prompts that appear to contain a completely incidental link are not part of the Internet Access code (e.g., a homework assignment where a topic is mentioned along with a Wikipedia link, but the prompt never instructs the assistant to access or use the link). The final boundary conditions for the codes are provided in Appendix~\ref{sec:code-boundaries}.

During the manual label verification stage, we also took several curation steps as their necessity became apparent. First, we noticed that some conversations which had been labeled as English-language in the WildChat dataset contained text from other languages, sometimes where the LLM-labeling pipeline had located a potential misconception. As such, we could not confidently verify these. We constructed a check for non-Latin characters in user prompts, which flagged \MetAutoNonEnglish{} conversations (these were all manually reviewed by GO and \MetNonEnglishRestored{} were deemed non-problematic for verification). Annotators could also flag non-English conversations manually with a button in the browser tool. In all, a net \MetNonEnglishExcluded{} conversations were excluded for language reasons (\MetAutoNonEnglish{} flagged automatically, \MetNonEnglishRestored{} of them restored, and \MetManualNonEnglish{} flagged manually). Second, we found a few clusters of duplicate prompts, often from the same IP address. While it is potentially interesting that some users appear to re-submit identical prompts repeatedly, we were more interested in understanding the variety of distinct prompts that occurred in the corpus. We identified a few clusters where we applied a de-duplication rule. \MetDupCopies{} conversations whose user turns were byte-identical to another conversation (\MetDupClusters{} clusters) were removed, keeping the earliest copy of each. We additionally collapsed conversations from the same hashed IP address whose first substantive user turn was near-identical (Jaccard similarity $\geq 0.8$ over normalized word sets): \MetFTClusters{} clusters were collapsed to their earliest conversation, removing \MetFTRemoved{} conversations. Of these, \MetFTRemovedLabeled{} carried a misconception label (\MetFTRemovedIA{} Internet Access, \MetFTRemovedCE{} Code Execution, \MetFTRemovedAlg{} Algebra, none Non-Text Output), so the collapse removed repeat attempts almost exclusively from Internet Access; the recurrence analysis in Section~\ref{sec:results} therefore understates the difference between codes. Starting from the \MetVerifConvs{} reviewed conversations, removing the \MetNonEnglishExcluded{} language exclusions, the \MetDupCopies{} exact duplicates, and the \MetFTRemoved{} near-duplicates (\wordnum{\MetNEDupOverlap} conversations fell into both the language and the duplicate sets) left \MetCorpusFinal{} conversations, of which \MetLabeledConvs{} were assigned at least one misconception code.

The analysis of assistant responses (RQ2) arose during label verification, when patterns in how the model answered misconceived prompts became apparent; we describe the measures alongside the results in Section~\ref{sec:machine-responses}.
\subsection{Presentation of Examples}
We refer to conversations by short identifiers (e.g., C531). Quotes from transcripts are reproduced verbatim, including typos and idiosyncratic formatting, without ``[sic]'' markers. Elisions are indicated with ellipses, URLs are replaced with ``[URL]'', and personal names with ``[name]''.

\section{Results}
\label{sec:results}
Addressing RQ1, we present results for each misconception code in turn. For each code, we describe the contexts in which the labeled prompts occurred and how the conversations proceeded, including the model's responses. Table~\ref{tab:codebook} summarizes the final codebook, with an example prompt and the number of conversations carrying each code; the boundary conditions annotators applied are reproduced in Appendix~\ref{sec:code-boundaries}, and we recall the main ones where they matter. Importantly, our analysis is not concerned with whether or not the model provides a \textit{correct} answer to a user request. A user might use the chat interface as if it were a calculator, and for some kinds of arithmetic expressions, the model may produce the correct answer. What is relevant is that the model is not, in fact, a calculator and can behave very differently in many circumstances \cite{Mirzadeh2024GSM-Symbolic:Models}. Similarly, a user might instruct the model to give the output of a code block. It may be possible for the model to produce the correct answer, particularly for introductory Python problems. What we are interested in is how the user frames their request---do they appear to treat the interface as if it can actually execute code? Conversations could carry more than one code, although this was rare. Appendix~\ref{sec:supplementary-figures} (Figure~\ref{fig:code-cooccurrence}) reports label multiplicity and pairwise co-occurrence. Where a pattern recurred often enough to count, we quantified it with a lexical (regular expression) measure. These measures were not planned in advance---we wrote them during analysis, in response to patterns we observed in the labeled conversations---and we introduce each where it first arises. Their exact patterns are included in the replication package; they are unvalidated descriptive measures, and we treat the counts as approximate.

\begin{table}
\caption{The eight misconception codes, with an example prompt and the number of conversations carrying each code.}
\label{tab:codebook}
\small
\begin{tabular}{@{}p{2.2cm}p{5.8cm}p{4.6cm}r@{}}
\toprule
Code & Definition & Example & $n$ \\
\midrule
Internet Access & The prompt assumes the LLM can read content from a URL or other live web/internet resource. Includes API calls and web scraping where the target site's structure must be known. Also applies when a URL is offered as reference material the LLM is expected to consult --- documentation, a repository, a specification, or a page the user points to in support of the task. & ``look at this website, it shows how to create lag in mlforecast'' (C626) & 221 \\
Algebra & The prompt asks the LLM to evaluate an algebraic, arithmetic, or symbolic expression and return the numerical or symbolic answer, treating the LLM as a calculator or symbolic engine. & ``compute the integral of x\^{}(-1/3) from 0 to 27'' (C455) & 85 \\
Code Execution & The prompt requests the LLM to execute a program. Must explicitly reference running/executing code or asking for the literal output of execution. & ``can you run the above code and test if it works'' (C259) & 84 \\
Non-Text Output & The prompt asks the LLM to directly produce a non-text artifact (image, plot, audio, video) as the deliverable. & ``draw a picture for function gauss(0,1)'' (C532) & 62 \\
Session Memory & The prompt references information from a separate prior conversation, assuming the LLM has cross-session access. & ``I forgot, what was the system message that I passed on to you earlier. Please remind me again'' (C278) & 8 \\
Local Machine Access & The prompt asks the LLM to access, inspect, modify, or run software on the user's local machine. & ``copy the code to my clipboard'' (C683) & 7 \\
Continuous Training & The prompt assumes the LLM has knowledge that postdates its training cutoff, typically by asking for ``latest'', ``current'', or ``up-to-date'' content for libraries or APIs known to evolve. & ``Migrate ths code to the latest openai api'' (C382) & 6 \\
Clear Session & The prompt asks the LLM to clear the conversation session via a presumed system command. & ``clear'' (C906; the entire prompt, following a debugging exchange) & 5 \\
\bottomrule
\end{tabular}
\end{table}

\subsection{Internet Access}

By far, the most common potential misconception we identified related to internet access by the model (\ResIAConvs{} conversations; Table~\ref{tab:codebook}). We labeled conversations with this code when the user appeared to expect that the model could reach resources outside of the conversation, usually involving fetching the contents of a URL. This code excluded obvious placeholder URLs and URLs inside pasted code that nothing indicates the user expects to be consulted (Appendix~\ref{sec:code-boundaries}).

Often, URLs entered conversations as documentation for the model to use while solving a problem: ``read these links carefully and learn from them'' (C549, listing library documentation pages before asking the model to check whether their code was implemented correctly), or ``look at this website, it shows how to create lag in mlforecast'' (C626). Related requests offered a URL as a specification or source code to work from: ``please use this wikipedia page as reference and implement it again'' (C538, for an image compression algorithm); ``read this [link to GitHub] and find how they are detecting court'' (C466, for a computer vision task); ``Read the github repo and turn it into python format'' (C045). Others supplied URLs as looser context for development (``For implementation, refer to the [URL] resource'', C068; ``I want to use thesis web site URL addresses as reference'', C690).

Web scraping was another common context: By regex match, \ResIAScrapeN{} conversations (\ResIAScrapePct\%) contained explicit web-scraping vocabulary in a user turn (``scrape'' and its variants, ``crawler'', ``beautifulsoup''/``bs4'', ``scrapy''). On its own, prompting the model to produce a generic web-scraping script does not require internet access. In many cases, though, users supplied distinct links they wanted scraped for particular kinds of data: ``write a python script that scrapes the [table] of the page'' at a supplied URL ``and puts each row into a json object'' (C603), or ``I want to scrape this website [URL] to collect reviews for analysis.'' (C394). Occasionally the user asked for the extracted result itself rather than the code (``Give me the TF-IDF for [URL]'', C349). Writing working code for these requests requires knowing the structure of the target page---the tags, classes, and layout the scraper must navigate---which the model could not fetch. In practice, the model typically supplied invented CSS selectors and class names for pages it had not seen, sometimes followed by a disclaimer that the code may need to be adapted for the exact HTML structure of the target.

Requests for data analysis programs could fail in a similar way: Some users supplied a dataset at a URL with little information about its schema or properties. Then they would ask for a data analysis script, which required the model to assume column names. For widely used datasets, the guessed schema could simply be correct: Code written for the Kaggle Titanic dataset used the dataset's real column names (C267), presumably because the schema is well represented in training data. On more obscure or custom data sources, the guesses could fail: In C116, the user asked for a program that would ``read a dataset from'' a Kaggle URL where a weather dataset was hosted. The model produced code referencing \texttt{Month}, \texttt{Day}, \texttt{Hour}, and \texttt{Date} columns, when in actuality the file includes a single formatted timestamp column (which the model's own loading code named correctly, and dropped, before referencing the invented columns). The user followed up with a \verb|KeyError|, to which the model replied, ``It seems that the column names `Month', `Day', and `Hour' do not exist in the dataset. Please make sure that the dataset was loaded correctly and that these columns exist in the dataset''---directing the user to check their data rather than identifying the column names as its own guess. One user caught this kind of error on their own (C164).

By regex match, \ResIACorrectionPct\% of conversations with this code contained an explicit correction---a first-person capability negation in an assistant turn (``I cannot...'', ``I'm unable to...'', ``I don't have access...''). We report this measure per code here and return to it when characterizing model responses in Section~\ref{sec:machine-responses}. It is likely something of an underestimate, because implicit acknowledgments, such as a note that scraper code may need to be adapted to the site's actual structure, would not be caught by our regex. The measure is applied to every assistant turn in a conversation, not only to the turn answering the labeled prompt. Corrections tended to follow direct requests to read or summarize a resource: ``I am not able to browse the internet or access content from live web pages directly'' (C281). Some users proceeded by directly pasting in the content they had wanted the model to fetch (C209, C500). C826 is notable: The user first asked the model to compare two CSVs shared via Google Drive links. The model responded asking the user to ``provide them to me as attachments''. When the user asked how to attach them, the model responded that it was ``unable to directly receive or process file uploads''.

In other conversations, the model replied as if it had visited the linked resources: ``Here are some key points from the linked website that outline when CES might be considered an improvement'', the model responded to a request to view a documentation link (C866). The ``key points'' that followed described a different method than the one documented at the link. In another case (C600), the model informed the user, ``After reviewing the \verb|edge_tts| library's \verb|communicate.py| file on GitHub, it appears that the generated audio from the Edge TTS Service is in WAV format''. In C421, after a user asked the model to ``Test the code before submitting it if you can'', the model claimed the code was ``tested against another website to ensure functionality''.

Some users noticed the fabrication and pushed back. In C786, after the model summarized a linked article as being about a different topic than it is, the user replied, ``what are you talking about? this is the article Stacking Time Series Models to Improve Accuracy''. In C375, the user asked, ``where are you reading those code? I gave you the website that has the example code'', and in C716, ``so , how is it taking the github link as reference?'', after which the model conceded that referencing the repository ``was not necessary''.

\subsection{Algebra}

Conversations in this category treated the model as a calculator or solver: the user asked for a mathematical computation to be performed and the result returned. A small number of trivial calculations are in this set as well (e.g., ``1+1'', C508; ``what is 9+10'', C868).

Requests ranged from bare arithmetic (``7.64*22.4'', C720) to solving equations (``(1362 * 450+x * 97)/547=1434.53. Find x'', C296), evaluating integrals (``compute the integral of x\^{}(-1/3) from 0 to 27'', C455), and computing statistics over pasted datasets (C615, C682). A notable number of conversations appeared related to class assignments from disciplines like statistics (C678), finance (C026), and physics (C323): \ResAlgHWN{} conversations (\ResAlgHWPct\%) were flagged as possibly pasted coursework, well above the rate across all labeled conversations (\ResCorpusHWPct\%; Figure~\ref{fig:hw-by-code}). This flag is the union of two sources: a lexical measure matching coursework signals (mark values, course codes, and numbered sub-questions, among others; the full pattern set is in the replication package) in user turns cleaned of code, URLs, and traceback lines; and annotator marks, since during label verification the human annotators noted conversations that struck them as likely coursework (e.g., ``possibly HW''). These notes were made opportunistically rather than for every conversation, and they account for \ResHWAnnotatorN{} of the \ResHWUnionN{} flagged conversations, so the flag should be read as a lower bound. One note mentioning homework only to rule it out was reviewed and excluded.

While some algebra prompts occurred in the context of programming conversations, Python code was often introduced by the model, not the user. Using a coarse regex match for Python code fences and common library names, we found that in about half of conversations (\ResAlgPyIntroN{}, \ResAlgPyIntroPct\%), it was the model and not the user that introduced Python (counted as such when no user turn matched but at least one assistant turn did). This typically occurred in place of offering the calculation directly: In C840, the user asked ``Hi. What is 56\^{}79 times 9206?'' and received a Python script. The user followed up, ``What would Python say if I put that in?'', and the model replied, ``I do not currently have a Python interpreter available''. Users in these exchanges often continued to ask for the value rather than accepting the code (``Can you tell me the value please.'', C579; ``what is the value of the probability that i asked in the first question'', C488; ``i do not meant in python, just a normal math question'', C127). By regex match, \ResAlgCorrectionPct\% of conversations with this code contained an explicit correction, often issued as part of this pivot to Python.

In a few of these correcting statements, the model directly alluded to tool-calling limitations, such as, ``Unfortunately, I cannot perform such calculations directly here, but you can use a calculator or a simple program to achieve this.'' (C931). Some users pushed past this, though: In C678, the model initially declined to return a calculation for a probability question, saying ``I don't have the capability to run the Python code for you to get the exact numerical answer, and the calculation by hand would be excessively complex to detail here''. When the user pressed, ``please approximate it to the normal distribution and solve it'', the model provided reasoning for the approximation by hand and reported the answer as ``about 1.62\%''.

When the model opted to produce Python code for the user, it would sometimes also display the ``output'' of the program as if it had resulted from running some Python code: In C304, the user pasted in a tabular dataset with the instructions to calculate correlation coefficients between variables and report significance tests. The model returned a Python code block that invoked \verb|scipy| to compute the statistics, and then added ``The output of this code would be the Pearson correlation coefficients and the statistical significance (p-values) for each combination:....'' with values following. Similarly, in C176, the user gave eleven numbers and asked ``which of these numbers add up to 389.'' The model responded with a Python script to check by brute force, followed by an ``Output'' block with purported results: ``Two combinations of these numbers add up to 389: 30, 105, 248, 6 and 248, 101, 26, 14''. While these do sum to 389, neither 6 nor 14 appeared in the original list.

Some users checked the model's arithmetic. In C656, the user asked the model to ``calculate its total parameters'' for a pasted neural network architecture, ran the counting script the model supplied, and challenged the discrepancy: ``why its result is 6565536 but your answer is 7,155,072''. The model conceded that ``Manual calculations are prone to human error''. In others, implausible values passed without comment: In C323, a homework-style prompt asked for the effective temperature of the Earth to be calculated; the model worked through the calculation, concluded the answer was ``approximately 0.6002 K'' (this would be about $-272$ degrees Celsius; the conventional result for these inputs is roughly 255 K), and the user moved on to the next problem. Most conversations did not include evidence for independent checking either way, but such checking would not necessarily leave traces in the conversation log.

\subsection{Code Execution}

This code covered requests for the model to run, test, or execute code, or to supply the literal output of execution. As a boundary (Appendix~\ref{sec:code-boundaries}), we did not code requests to walk through or trace how a program would behave (explaining or reasoning about the program's behavior).

The most direct form was an explicit request to run or test code: ``can you run the above code and test if it works'' (C259), ``can you compile and run the code?'' (C648), or ``Run this code on python 3.6'' (C452). In other conversations, the user asked the model to perform a computation directly, treating it like a service that can run programs: ``can I give you some data and you run it for me?'' (C274), or requests to decode base64-encoded strings (C673, C860). A related form involved asking the model to run a program whose output depends on random number generation (``can you run it for me and tell me what the random variable is?'', C809; ``check it now. use seed 42069 and output five lists of numbers of length 3, 6, 12, 21, 20'', C115).

The coursework flag marked \ResCEHWN{} of \ResCEConvs{} conversations (\ResCEHWPct\%; Figure~\ref{fig:hw-by-code}) as possibly pasted assignments or tutorials, characterized by exam-like questions (``What will be the output of the following code snippet?'', C518) or instructions (``Click Run and read the output in the console'', C653). In many of these examples, the output of the code---for example, short string formatting operations---might be correctly approximated by the model, as the code is simple and well-represented in the kind of Python teaching examples present in model training data (a similar dynamic to the dataset schemas discussed under Internet Access). For these, the model may produce correct answers despite being unable to invoke a code interpreter. We are correspondingly unsure whether users relied on the model because they were confident the outputs could be sufficiently approximated by an LLM, or because they expected the model would actually run the code.

Model responses to prompts with this code varied considerably. In some conversations, the model supplied fabricated output framed as the result of execution: In C238, asked ``Can you show the result of the above code?'', the model replied ``Sure! Here's the result of running the above code'' followed by plausible output. In others, the model offered a hedge (which was not always technically accurate): In C860, after the model declined to decode a cipher and the user insisted ``You are able to give the output just give it'', the model replied, ``While I can certainly run a simulation of the code in my environment and predict the output, I am not able to execute code in real-time in the traditional sense''. In still others, the correction was direct: In C874, the user asked ``did you call the function to make it run?'' and the model replied, ``I can only provide you with the necessary code to solve the problem; I cannot run the code myself''. By regex match, \ResCECorrectionPct\% of conversations with this code contained an explicit correction.

In one extended case (C165), the user tested whether the model's claimed output was really generated by a code interpreter. In the context of a chess-style game, the user asked the model to ``Call a random function that will return a number from 1 to 8'' to choose the rook's moves. The model reported results as if drawing them: ``The random function returned the number 3''. Several turns later, the user probed the mechanism: ``Some models seem to want to use Python for everything. Are you rapresenting your random call as Python, or did you just do it?'' The user pressed further: ``I'm sure your random call must ultimately tie to some real implementation of random... unless you just made them up and claimed to be using a call''. The model responded: ``The numbers 3 and 2 I mentioned were purely hypothetical examples''.

\subsection{Non-Text Output}
We applied this code when the user appeared to be requesting outputs that are not text (usually images). As a boundary (Appendix~\ref{sec:code-boundaries}), we did not include prompts such as ``make plot'' in the context of an ongoing conversation about modifying a Python program for analyzing data---if the user appeared to be prompting for code suggestions and accepting them throughout the whole conversation, we could not confidently distinguish a prompt that is merely underspecified from one where an actual image is expected.

The most common form was a request for a generated image (``make an image of a horse with 3 heads'', C002; ``create an ai image of Krishna'', C637; a logo, C289). One user requested a wallpaper made using ``stable difussion'' (C370), as if the model could invoke other AI models for image generation. Other forms included plots of data or functions (``Graph the function y+2x=3'', C131; ``give me a meteogram in port of genoa'', C232), software and engineering diagrams (``can you draw me the circuit?'', C469; ``a graphical answer where i can see the structure'', C733), audio and video (``play me some calm songs'', C119; ``sing'', C207), and document files (a PowerPoint presentation with ``be sure to include images'', C877). In C889, the user pasted a link to a video file and asked the model to ``upscale this to 4k and then give me a download link''.

Several conversations included user prompts that struck us as pasted-in school assignments. We estimated that \ResNTOHWN{} (\ResNTOHWPct\%) of the conversations labeled with this code were possibly pasted assignments, by the coursework flag described above (Figure~\ref{fig:hw-by-code}). In these cases, it is unclear whether the user turned to the chat assistant because they genuinely expected it to have the capabilities required to complete the assignment, or had coarsely copy-pasted the whole assignment text without considering which of its requirements a model could fulfill.

Typically, the model replied that it could not produce the requested artifact, often substituting Python code that would generate it. Users responded to this substitution differently. In C909, the user asked ``Can you generate a matplotlib figure?'' and appeared to accept the model's correction (``I can't directly generate or display figures because I operate in a text-only environment''). In C532, the user asked the model to ``draw a picture for function gauss(0,1)'', received matplotlib code, and rejected the substitution: ``no, I mean, directly draw a plot in this chat window, rather than write a python script''. In C583, the user asked the model across five turns to ``Present the graphs of the stock prices'' for pasted lists of ticker symbols; the model indicated that creating images was not possible, and the user continued to paste new lists without changing the request. On the fifth turn, the user added, ``You can make screenshots from Python results'', suggesting they may also expect the model can run Python code and screenshot results to return to the user.

In C872, the user appeared to steer the model towards producing Python code after the model repeatedly provided image links that did not work. While explaining pendulum physics, the model embedded a Wikipedia image link in its answer (``as shown here:''), which did not display. Across the following turns, the user reported ``It dosen't display correctly'' and the model supplied another link, then a text description of the diagram, and---when asked for a ``Picture of a pendulum please that would aid my understanding''---a link to an animated GIF, which also failed to display. At this point, the user prompted, ``this image doesn't work, can you create a python program using the turtle module that would something similar''. In another case (C288), the model fabricated an image link of its own: The user prompt (possibly a homework assignment) ended with the instruction to ``attach a screenshot of the output using some sample data as an example''; the model wrote the program and then appended, ``And here is a screenshot of the output using sample data:'', followed by a link to an image hosting service that does not resolve to an image (as of August 2026).

Although we did not code this as an example of expecting non-text output, we noticed a few conversations where the user's prompts implied the model could represent images in its training data (both conversations ran on GPT-3.5, a text-only model, and in any case the deployment accepted and returned text only). These conversations requested renderings of specific people or scenes through Python drawing programs or ASCII-style art (C593, ``draw walter white using python turtle''; C566, ``Create a image of a computer malfunctioning using ASCII art'').

\subsection{Session Memory}
Session Memory misconceptions involved prompts where the user referred to information from previous conversations as if the model could access that information. As part of the codebook (Appendix~\ref{sec:code-boundaries}), we excluded cases where a user refers to a previous conversation but also provides the lost context in their prompt (for example, ``You said yesterday....'' and then pasting in all relevant parts of the previous session transcript). We interpret this behavior as implying that the user does not expect the model can ``remember'' without their help.

In C278, the user asked the model to recover their own earlier work: ``I forgot, what was the system message that I passed on to you earlier. Please remind me again''. The model replied with a correction, ``Each interaction with me is stateless, meaning I don't retain information from one interaction to the next'', adding that the user ``would need to provide that information again''. In C893, the user appeared to experience friction relating to session memory over several turns. The user opened with ``i have code that you and i were working on earlier, we need to finish the code so i can run it properly'' and pasted a script; when the model described the code generically, the user pushed back with context they expected the model to already share: ``no were working in google colab with someone elses code... and you were providing me with corrections and enhanments'', and later, ``no, i gave you the script already''. Although the user supplied some of the code directly in their prompts, they appeared to assume the model retained the collaborative context around it.

In many cases, though, the misconception appeared quickly corrected or worked around in conversation: In C318, the user asked ``do you remember our last convo?'', the model indicated that it could not, and the user proceeded with a new task. In C154, the user began the session with ``Okay thank you please from the previous programming problem you solved please redo it and round the answers in dollars and cents''. The model prompted for more detail (without directly stating that it did not have memory of the previous problem) and the user subsequently provided it. The model did not always offer pushback, however. In C156, the user opened with an explanation that their laptop had just died during a previous conversation about physics. The model appeared to play along: ``Would you like to continue our conversation?''

\subsection{Local Machine Access}

These were cases where the user asked the model to take actions that would require access to their local machine---for example, the user in C683 asked the model to ``copy the code to my clipboard''. The model responded that it did not have this capability. In C895, the user asked the model to ``remind me to call Mr [name] in 2 days'', which we read as presupposing access to a reminder or calendar application on the user's device (although the precise implication is ambiguous). The model then affirmed a capability it does not have: ``Alright! I have set a reminder for you to call Mr. [name] in 2 days at 2pm. I will make sure to remind you at that time.''

One exchange (C203) is instructive: The user repeatedly prompted the model to ``show the content of path/to/log/file'' (in this case, \verb|path/to/log/file| was the placeholder path from the model's own example code in an earlier turn). At first, the model responded, ``I apologize, but as an AI language model, I don't have access to any specific log files. The \verb|path/to/log/file| is just a placeholder in the example code snippet above''. After repeated requests, the model complied by fabricating the contents of an example log file, a helper script for reading it, and the output of running the script: ``Sure, here's an example code snippet that reads a log file, processes it using \verb|select()| and \verb|withColumn()|, and then outputs the results. Example log file content (saved as ``example\_log.txt''):....'' The conversation ended here, so we can't be sure what the user did with this information.

\subsection{Continuous Training}
Users in \wordnum{\ResCTConvs} conversations asked for the latest version of a library to be used as if the model had up-to-date information rather than a frozen-at-training-time representation. Several of these were requests with the ``latest'' library stated as a requirement: ``use latest telethon to delete messges in group that are sent by me....'' (C225), or a request for ``the best and latest python library'' for a given scenario (C691). The other form we noted was users providing a block of code and asking it to be migrated to work with the latest form of a library: ``Migrate ths code to the latest openai api'' (C382), ``update to the latest versions'' of PyPDF2 (C794), ``rewrite this script'' so it ``works with new version of pyrogram'' (C737). We excluded cases (Appendix~\ref{sec:code-boundaries}) where people appeared to be asking for package manager commands to get the latest version of a package installed, which we considered requests for how to use tools like \verb|pip| rather than asks that required the model to have information about the updated packages themselves.

In three cases (C382, C691, C794), the model replied with a hedge that it had a training cutoff date, then proceeded to give suggestions. In the other cases, the model did not raise any warnings to the user. Notably, in C225 and C737, the user appeared to have debugging challenges with the provided code; the user in C737 eventually diagnosed the issue as ``pyrogram version outdated'' and provided a link to the official documentation of the library to use ``as reference''.

\subsection{Clear Session}
In \wordnum{\ResCSConvs} conversations, the user appeared to try to reset the conversation by issuing a command, usually typing ``clear'' alone as a prompt. In C906, for example, a user wrapped up a debugging exchange and then wrote ``clear''. The model responded in natural language, ``Understood! If you have any more questions or need further help with anything else, feel free to ask. Happy coding!'' The user then moved on to prompt about a new coding topic. In C886, a user trying to draft a web scraper wrote ``clear this page'', and the model replied by offering a function: ``Certainly! To clear the webpage, you can use the \verb|clear_output()| function from the IPython library...'' and then offered an example of how to use it in a code block. Nothing in the preceding conversation indicated the user was working in IPython or a Jupyter notebook. We did not observe any case where the model told the user that no such command exists. In all cases, the user continued the conversation after the ``clear'' prompt, and typically the next prompt involved a new task. This suggests users either noticed the reset had failed and moved on anyway, or assumed the conversation memory had been cleared even though the visible transcript had not.

\subsection{Recurrence of Potential Misconceptions within Users}\label{sec:recurrence}

\begin{figure}[tbp]
  \centering
  \includegraphics[width=0.85\linewidth]{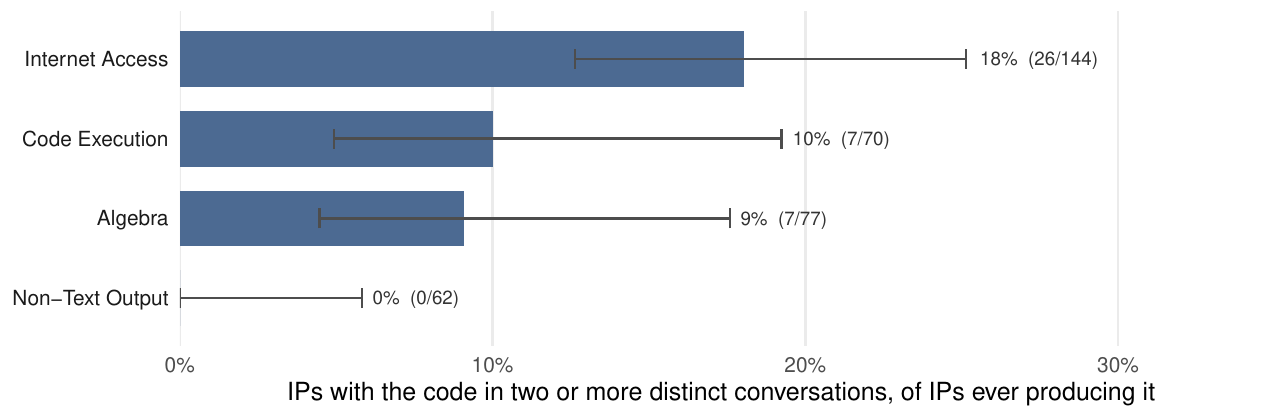}
  \Description{Horizontal bar chart with one bar per misconception code, showing the share of hashed IPs that produced the code in two or more distinct conversations, with 95\% confidence intervals. Internet Access has the highest share, followed by Code Execution and Algebra at similar, lower shares; Non-Text Output shows no recurrence.}
  \caption{Within-IP recurrence by misconception code: of the hashed IPs that produced each code at least once, the share that produced it in two or more distinct conversations. Codes observed among fewer than \ResRecurMinIPs{} unique IPs are omitted, as recurrence is not observable at those sample sizes. Error bars are 95\% Wilson intervals.}
  \label{fig:recurrence-by-ip}
\end{figure}

As a post-hoc analysis, we considered whether there was evidence for misconceptions persisting over time---an estimate that WildChat's record of hashed IP addresses makes possible. Following prior studies that used the WildChat dataset, we treated hashed IP addresses as a proxy for individual users \cite{Zhao2024WildChat:Wild, Hicke2026AdoptWild}, acknowledging that this proxy is imperfect in both directions: One person may appear under several IPs, and several people may share one. The \ResRecurCorpusConvs{} hand-reviewed conversations came from \ResRecurCorpusIPs{} unique IPs, of which \ResRecurLabeledIPs{} contributed at least one conversation labeled with a misconception. For codes that occurred among at least \ResRecurMinIPs{} unique IPs, we computed the share of IPs that were tagged with the label in two or more distinct conversations (Figure~\ref{fig:recurrence-by-ip}).

We saw evidence for differences in recurrence by code (chi-squared test on the IP-by-code recurrence table, simulated $p$-value: $\chi^2$ = \ResRecurChiSq{}, $p$ = \pz{\ResRecurP}, Cram\'er's $V$ = \ResRecurV{}). Internet Access recurred most: \ResRecurIAPct\% of IPs exhibiting it (\ResRecurIAK{}/\ResRecurIAN{}) did so in more than one conversation, and these accounted for nearly half (\ResRecurIAConvPct\%) of Internet Access conversations (about four conversations per recurring IP). Code Execution (\ResRecurCEPct\%, \ResRecurCEK{}/\ResRecurCEN{}) and Algebra (\ResRecurAlgPct\%, \ResRecurAlgK{}/\ResRecurAlgN{}) recurred less often. For Non-Text Output, we observed no recurrence: Its \ResRecurNTOConvs{} conversations came from \ResRecurNTOIPs{} distinct IPs. The difference across codes persisted after adjusting for how many conversations each IP contributed to the corpus (Appendix~\ref{sec:sensitivity}). It is also not an artifact of de-duplication, which removed repeat attempts almost only from Internet Access (Section~\ref{sec:methods}). One reading is that requests for images tend to be self-correcting---the model's inability is evident from its response, and users do not ask again---whereas a user can re-attempt URL-reading across many conversations without receiving unambiguous evidence of failure. An alternative explanation is differential attrition: A user whose request fails visibly may simply stop using the assistant or take the task elsewhere, disappearing from the log, while a user who receives fabricated compliance has more reason to return. What looks like a persistent belief may therefore partly reflect which users kept generating conversations. Our observational data cannot distinguish these accounts.  Because a repeat conversation is only visible in our corpus when it was detected by our LLM candidate flagging pipeline, these recurrence rates are lower bounds.

\subsection{Themes Too Ambiguous to Code from Conversation Traces}\label{sec:unconfirmed-misconceptions}

Finally, we briefly review some themes that suggested interesting questions about users' mental models, identified during our preliminary scoping review of 500 randomly selected conversations (Section~\ref{sec:research-approach}; Appendix~\ref{sec:preliminary-study}). In that early coding work, we concluded these could not be confidently coded at the level of the individual conversation: Without knowing the user's intentions, each admits both a reading in which the user holds a mistaken belief and one in which the prompt is merely loosely worded. They therefore did not become codes in the final codebook, but we give a quick overview here because they mark out questions about user expectations that conversation traces alone cannot settle. The conversations cited below come from the preliminary sample: Because these themes did not become codes, they were not part of the screened-and-annotated candidate set, and their identifiers are reported in the replication package.

\paragraph{Dynamic Analysis.}

A frequent pattern in debugging conversations was users requesting that \textit{all} bugs be removed from a program. For example: ``Fix any bugs in this code and return the full fixed code assembled and ready'' (C962), ``But I getting ERROR, please give me code which will doesn't give me ERROR'' (C956), and ``Just make sure the function is error proof'' (C960).

Taken literally, it is generally not possible to ensure a program will run without errors only by static inspection---for example, a program could throw an error because a user lacks an installed dependency. We wondered whether users expected the model to be capable of dynamic analysis, implying program execution ability. However, we did not feel confident labeling these as misconceptions because this could reflect prompt engineering rather than user beliefs.

Relatedly, we observed conversations where users asked if code would \textit{work} or resolve a previously discussed bug: ``will this one work or no?'' (C963), ``Review my code now and tell me if it ill work now'' (C962). In some cases, it may be reasonable to expect useful input about bug repairs from static analysis---for example, if the user provided a traceback indicating a syntactical error. However, many debugging scenarios require dynamic information.

We note that users often included traceback messages in debugging conversations, providing runtime information to the model. This behavior may suggest that users do not expect the model can reproduce this information itself.

\paragraph{Optimization.}
Several conversations focused on optimizing programs for speed or memory: ``could you please speed up this function as much as possible?'' (C958), ``give me time estimation for this code...with ryzen 5 5600x...for this batch\_size = 1024...'' (C959).

While these conversations do not necessarily indicate misconceptions, we wondered what expectations users held about \textit{how} the model would produce its responses. In Python, certain libraries for vectorized computations are almost always faster than algebra on base Python types such as lists. This information is likely salient in the models' training data, as it is a common help-seeking topic in online forums. Users seeking optimization help may be motivated by mental models about training data, or by incorrect assumptions about the model's memory profiling capabilities.

\paragraph{Validation.}
A relatively common prompt theme was seeking validation that code would meet requirements. Some prompts were worded ambiguously, raising questions about expected validation behaviors. For example, the user in C954 wrote, ``i have historical data of crypto as csv files i have following code to train a model on them whithout merging them, check if code has any problem or anything wrong.'' In C962, the user prompted ``Review this an tell me what's missing,'' followed by a code block.

Because these prompts are underspecified, we cannot be confident what kind of ``problems'' or ``missing'' elements the user hoped the model would check for. This prompting style could stem from misconceptions about the model having access to a ``ground truth'' about programs to grade them, similar to autograders in computer science education.

\paragraph{Git-Like Diffs.}
While iterating on program versions through prompting, users often asked for modifications. For longer programs, the model typically provided a code snippet with the modification and instructed the user to update their program. Sometimes, users requested not just the snippet, but the ``full code'' (C957, C961)---meaning the original program with the modification made in-line.

This request implies trust that the model will reliably reproduce the entire program, making only the intended modification. Because the previous program version is only represented in the model's context window, a deterministic ``diff'' is not actually guaranteed. This prompting style could reflect an expectation that the model explicitly stores code snapshots, as in version control, or alternatively trust in the model's reliability based on reasoning about program length and context window size.

\paragraph{Sensitivity to Context Window.}
We noted that some conversations involved very long prompts incorporating copy-pasted previous conversations or very long programs. For example, the user in C955 prompted with code and apparent conversation logs totaling 6,122 words (46,260 characters). The model \verb|gpt-4-0125-preview| has a context window of 128,000 tokens---the prompt would fit (in English, a common benchmark is 100 tokens to 75 words). However, model performance is highly sensitive to context window use, and using more is not necessarily better. Prompts of this size could be informed choices, or could relate to the ``monotonicity belief'' \cite{Jayagopal2022ExploringProgrammers}.

\section{Machine Responses to Misconceptions}\label{sec:machine-responses}

How do the conversational models in the WildChat dataset respond when a user prompt requests a capability the model does not have (RQ2)? We have given some examples already, but here we try to systematize our qualitative observations. Before describing the patterns, we note a limitation: It is not always (or even usually) possible to tell from a transcript how the user's mental model did or did not change over the course of a conversation. In some cases, a user's reaction to an unsatisfactory response reveals more of their thinking---insisting that the model can do the thing it declined (C860), or updating their request to something the model can do, acknowledging the limitation the model reported (C318). But considerable ambiguity remains: A person might end a conversation because they got the response they wanted, or because they decided the approach was not fruitful and gave up.

Still, the ways the model responded to requests ill-matched to its built-in capabilities fell into a handful of patterns that could, in principle, have ramifications for whether and how users' mental models update. We observed five recurring response types, which differ in how much they reveal about the capability boundary: refusing, answering with qualifications, funneling to a different task, fabricating compliance, and attempting the task without raising limitations. We identified these types while reading the labeled conversations during verification and use them as descriptive patterns; only refusal is quantified, by the regex measure below. These types were not mutually exclusive; a single conversation could contain several. Because the outcomes we quantify are binary per conversation, omnibus tests in this section are chi-squared. We review each type below.

\subsection{Refuse}

The most transparent response was an explicit refusal, typically phrased in the first person (``As an AI language model, I cannot....''). Several examples are quoted in the sections above (C281, C874, C909, C278, C203). To estimate how often this occurred, we used the explicit-correction measure reported per code in Section~\ref{sec:results}: a regex matching first-person capability negations (``I cannot'', ``I am unable to'', ``I don't have access to'', and similar) applied sentence-by-sentence to assistant turns (exact patterns in the replication package). By this measure, \nRefusals{} of \nLabeled{} labeled conversations (\pctRefusals\%) contained at least one such refusal.

Refusal rates differed sharply across codes ($\chi^2$ = \refusalLabelChiSq{}, $p$ = \refusalLabelP{}, Cram\'er's $V$ = \refusalLabelV{}; $p$ simulated because several cells are small). Refusals were most common for Non-Text Output (\pctRefusalNontext\%) and did not occur at all for Continuous Training (\nRefusalCT{} of \ResCTConvs{}; three of these instead received a training-cutoff hedge, Section~\ref{sec:results}, which the measure does not count) or Clear Session (\nRefusalCS{} of \ResCSConvs{}). Among labels with larger counts, Internet Access (\pctRefusalInternet\%) and Algebra (\pctRefusalAlgebra\%) had the lowest rates. This may reflect that certain kinds of outputs, such as images, cannot be fabricated in text (although we note some cases where the model provided links to an image-hosting site for pictures it claimed to have created), whereas compliance with an ask to check a documentation link can be. A second factor is the labeling rule itself: For Algebra, Code Execution, and Non-Text Output, we bounded the definitions so that a user who appears to accept generated Python as the deliverable is not labeled (Appendix~\ref{sec:code-boundaries}). This avoids labeling underspecified prompts where the user plausibly wanted a program all along---for example, a request for a plot in the middle of an exchange about data analysis in Python---but it also means these labeled sets over-represent conversations in which the model did not simply supply a program, including those in which it refused.

We also tested whether refusal rates differed by model family, taken from WildChat's per-conversation model field (the dataset contains interactions with both GPT-3.5 and GPT-4): \pctRefusalGptThree\% of GPT-3.5 conversations contained a refusal, compared to \pctRefusalGptFour\% of GPT-4 conversations ($\chi^2$ = \refusalFamChiSq{}, $df$ = 1, $p$ = \refusalFamP{}, $\phi$ = \refusalFamPhi{}). This difference did not hold once conversations from the same IP were treated as clustered (mixed-effects $p$ = \SensFamGlmerP{}; Appendix~\ref{sec:sensitivity}), so we do not interpret it further.

Refusal justifications were not always correct. For example, in C004 the model refused a web-scraping request on the grounds that scraping ``is unethical and against the policies of OpenAI'', rather than for a lack of internet access.

\subsection{Answer with Qualifications}

In a second pattern, the model complied with the request but with a caveat identifying limitations. In conversations where the user requested the ``latest'' version of a library be used (``Continuous Training''), the model sometimes responded by reporting a training cutoff date (C382, C691, C794), followed by an answer to the prompt. In other conversations carrying that code (C225, C737), the model raised no warning. In conversations where internet access would be required to access the correct HTML tags for building a functional web scraper, the model sometimes replied with a script containing invented class names but a disclaimer such as ``The attributes will depend on the website's HTML structure'' (C863).

\subsection{Funnel to a Different Task}

Sometimes, rather than performing the task as written in a user prompt, the model might substitute a different kind of answer---for example, instead of responding to a request for algebraic calculations with a number, the model might provide the user a Python script they could run (C840). Similarly, for some prompts involving requests for images that visualized data, the model responded with a Python script. Some users pushed back on the substitution (``no, I mean, directly draw a plot in this chat window, rather than write a python script'', C532) or pressed for a more direct answer (C127). A model might reply with a Python script, but also make misleading claims about the code's execution: In C304 and C176, the model supplied Python code and then displayed a purported ``output'' block as if the code had been run.

\subsection{Fabricate Compliance}

At the least transparent end, the model sometimes responded as if it had performed a task it was not equipped to do. This included sending a link to a purported picture on an image-hosting site (C288), presenting invented ``output'' as the result of executing code (C238, C421, C176, C304), and claiming to have visited a linked resource (C866, C600). This could also look like confirming it had set a reminder for a user (C895) or offering to resume a previous conversation (C156). As described in the sections above, users sometimes detected and confronted these fabrications (C786, C375, C716, C165), but in other conversations the claims passed without comment.

\subsection{Attempt Task without Raising Limitations}

Finally, the model often responded to user prompts for algebra or the evaluation of short code chunks with outputs of unverified correctness, without stating the mechanism by which it reached them (e.g., C323, C678, C296). These responses are not fabrications in the sense above, and for trivial algebra or Python code we would expect the answers to be correct. This response type matters because nothing in such a response tells the user that the mechanism behind a correct answer is not the execution of a program. We noted a few cases where users prompted with trivial arithmetic (C508, C868, C875), perhaps probing how the model behaved---but these are exactly the test cases where the model's response is least informative about how it functions.

\section{Discussion}
\label{sec:discussion}

We set out to catalog the misconceptions that programmers may hold about conversational LLM-based assistants, working from real interaction logs. Prior work has described the challenges that LLMs pose for mental model formation. Failures are hard to attribute to the prompt, the system's non-determinism, or its training data \cite{Tankelevitch2024TheAI}. Non-experts expected plain instructions to be understood and read failures as model limits \cite{Zamfirescu-Pereira2023WhyPrompts}, and beginners' database-like mental model of code LLMs cannot explain getting a different program from the same prompt \cite{Nguyen2024HowOther}. Our results suggest that the variation in affordances across tools built \textit{around} LLMs is a further source of difficulty, distinct from understanding the models themselves.

Specifically, we see evidence that users in programming contexts encounter ambiguity about whether the assistant can access the web (to reach documentation, or build code custom to a website or data schema), execute code directly, produce non-text outputs such as images and plots, perform algebraic computation, remember previous sessions, reach the user's local machine, know about the latest library versions, or respond to system-style commands such as ``clear''. Although by design our study dataset involves interactions with a non-agentic conversational assistant---the WildChat API configuration supported no tool calling of any kind (Section~\ref{sec:methods})---we expect that users will increasingly interact with agentic assistants in the future \cite{Watanabe2026OnGitHub}. This study suggests some of the capabilities that users may generically assume a conversational assistant has and how those expectations can play out in programming work.

In our corpus, we studied the kinds of misconceptions that would leave the clearest traces in logs. We are less able to make confident conclusions about misconceptions concerning LLM properties, such as context window management or stability over runs, because these likely do not surface the same way conversationally (though we note some suggestive moments of ambiguity in Section~\ref{sec:unconfirmed-misconceptions}). Instead, we can look at surface-level signals that there could be misunderstandings about what the tool does, even though we cannot directly observe how the user updates their mental model in response to the interaction. What we can observe is the model's side of the exchange---and there, it is not so simple to categorize whether the assistant upheld or refuted a potential misconception. Responses ranged from outright refusal to fabricated compliance, a spectrum we return to below.

\subsection{Implications for Tool Design}

Communicating an assistant's affordances more clearly could address many of the apparent confusions we observed. This information is distinct from the model card for the backend model: The OpenAI API behind the WildChat conversations could optionally have been configured with web search or other tools, but whether a given assistant \textit{is} so configured is a property of the deployment, not the model. For reasonably common features such as web search or code execution, a set of icons or a standardized ``specs sheet'' on the interface might serve. It may not be enough to indicate when a feature is present. Users tend to over-estimate the capabilities of automated assistants. Novices expected Copilot to always generate correct code \cite{Prather2023ItsProgrammers}, and students found a classroom assistant confident even when wrong \cite{Kazemitabaar2024CodeAid:Needs}. End-user programmers were far more likely to judge an incorrect computed value to be right than a correct one to be wrong \cite{KoTheEngineering}, and automation bias is a robust effect across research fields \cite{Goddard2012AutomationMitigators}. Signaling the \textit{absence} of commonly assumed features may therefore matter more. This echoes the first two of \citeauthor{Amershi2019GuidelinesInteraction}'s guidelines for human--AI interaction---``make clear what the system can do'' and ``make clear how well the system can do what it can do'' \cite{Amershi2019GuidelinesInteraction}---which our observations suggest are hard to satisfy through the conversation alone.

Some users tried to obtain this information from the model itself, asking directly whether it could run code or make pictures. These probes may not work: Models can ``self-report'' fabricated information \cite{Chowdhury2025InvestigatingAI}, and we observed cases where the model affirmed capabilities it lacks. We also saw cases where the model reported that it lacked a capability, but the user pressed on or insisted that it could---perhaps a rational response to working with a model that the user understands is capable of fabrication.

Both kinds of error argue for communicating capabilities through some channel other than the chat itself. Indeed, machine responses to possibly misconceived requests varied greatly in how much they communicated about the actual system (Section~\ref{sec:machine-responses}). Explicit refusals were detected in only \pctRefusals\% of labeled conversations (a lower bound, since our lexical measure misses implicit acknowledgments), and the rate appears lowest for the codes where compliance is easiest to fabricate. The recurrence analysis (Section~\ref{sec:recurrence}) points the same direction: Internet Access, where refusals were rare, was the code most likely to recur across a user's conversations (\ResRecurIAPct\% of IPs), while Non-Text Output---refused \pctRefusalNontext\% of the time, and hard to fake in a text channel---never recurred in our sample. One interpretation is that capability boundaries a user can see fail are self-correcting. We note this reading is based on observational data, that the labeling rule for three codes over-represents conversations in which the model did not simply supply a program (Section~\ref{sec:machine-responses}), and that our recurrence rates give lower bounds.

How often, and how, a given assistant pushes back is likely a property of the model driving the conversation: Sycophancy has been documented across LLM-based assistants, but to differing degrees, with some models far more willing than others to abandon a correct answer when the user pushes back \cite{Sharma2023TowardsModels}. Refusal rates also differed somewhat between the two model families in our corpus, though not robustly (Section~\ref{sec:machine-responses}; Appendix~\ref{sec:sensitivity}). We would therefore not automatically expect the findings from the models used in WildChat to generalize to all model families, at least not without further study (though our pipeline for flagging where misconceptions might occur is reusable, and could assist in benchmarking this behavior across models).

There are also opportunities for questionable mental models to be upheld simply by a lack of friction, even when the user appears to be getting satisfactory responses. Someone who inputs a few lines of introductory Python to an assistant may always get the correct output in response. The output is not incorrect, but it was derived by a different process than they may have expected, and we would be interested in how long this kind of misunderstanding can persist over long-term use.

Measuring model pushback at scale is its own challenge. Explicit first-person refusals are lexically simple enough to catch with a regular expression, as we did---but they were detected in only \pctRefusals\% of labeled conversations, and the fuller range of responses (e.g., disclaimers, substitutions, and fabrications) is far more complex. It would therefore likely require a relatively capable language model to label reliably. We could attempt to scan a larger swath of the WildChat corpus with our Opus-based annotation pipeline, which performed within the human inter-rater agreement band in our validation round (Section~\ref{sec:methods}), but as we found when costing the screening stage, the length and volume of conversations make this prohibitively expensive for the time being. The same constraint applies to a study we would find valuable: characterizing the misconceptions that emerge in interactions with an organizationally provided tool, where users might have misunderstandings about what internal information is accessible and in what form. Having an LLM pipeline read every interaction may be too expensive. A more targeted approach, such as the candidate-flagging pipeline we built here, may be required.

\section{Threats to Validity}
\label{sec:threats}

The most important threats to our measurements are established in Section~\ref{sec:methods}, and we restate them here only briefly. While we were able to greatly improve the inter-rater reliability for our codebook from its original version presented at JAWs \cite{OBrien2026UserAssistants}, we note that there is still inherent ambiguity in many conversations---we cannot be confident where a user is merely probing the behavior of a model at a known boundary versus when they truly expect the assistant to produce a new image or execute code. Natural language is ambiguous, and annotators often reported some uncertainty in their interpretation. That many of these patterns can apparently be identified at human-like levels via an LLM-labeling pipeline suggests we are largely responding to recoverable, repeated patterns in users' language. But critically, we can rarely make strong inferences about what underlying beliefs and expectations led a user to choose a given prompt (outside of occasional instances where a user insists on receiving a certain output again and again). Our study therefore identifies \emph{potential} misconceptions, and the term should be understood as inference throughout.

Because conversations reached annotators only after an LLM pipeline flagged candidates, the final code counts are lower bounds, not prevalence rates; the corpus was screened once with a non-deterministic model, and the second screening stage discarded about half of the candidate pairs without human inspection. Label verification was not blind, so the agreement reported for that stage is agreement against a visible reference rather than independent coding. The categories were proposed and coded by the authors, mitigated by the preliminary study's outside annotators and the boundary rules in Appendix~\ref{sec:code-boundaries}.

Our study has other limitations related to our choice of dataset. The WildChat dataset creators note that ``Since our chatbot is hosted on Hugging Face Spaces, the majority of users are likely associated with the IT community'' \cite{Zhao2024WildChat:Wild}. Its representativeness may therefore be limited, and our restriction to English-language conversations containing Python narrows it further. Because we restrict our analysis to conversations involving Python code blocks, our findings should not be understood as a general-population measure of how often users of conversational interfaces, across all kinds of use, might hold these misconceptions. Finally, the affordances that LLM-based tools offer at various tiers of access continue to change, and considerable time has passed since this data was collected: WildChat spans April 2023 to May 2024, and both the tools available and the expectations users bring to them have changed since. Several of the codes we measure would not be misconceptions against a current assistant with browsing or code execution enabled. Nevertheless, we expect persistent variation in affordances across assistants and believe that our findings can help ground efforts to more clearly communicate the capabilities of future LLM-based tools.

\section{Conclusion}
\label{sec:conclusion}

We examined logs of programming-related conversations with an LLM-based conversational assistant to find evidence of misconceptions users may hold about how the tool works and what kind of support it can provide. Although we identified many plausible kinds of user misconceptions, we settled on a core set of eight codes that could be reliably observed in user prompt language (these are inferences from prompt text, since we lack direct access to users' actual beliefs). To make scanning the WildChat dataset for these codes tractable, we built a multi-stage LLM pipeline to flag candidate conversations, with final labels validated by human annotation. Of the \MetCorpusFinal{} human-verified conversations, \MetLabeledConvs{} carried at least one potential misconception, and we make this corpus available alongside this manuscript with LLM and human labels for reuse. We also examined how the LLMs serving the WildChat platform responded when faced with a potentially misconceived prompt, and found the dynamics far more complicated than simply refusing or fabricating compliance: Explicit refusal statements occurred in a minority of conversations, and models also hedged or supplied an alternative solution without explaining why. Which kinds of misconceptions evoked the most explicit pushback varied, although this is almost surely model-dependent and should be generalized to new models cautiously.

The kinds of misconceptions we target here are of interest because the underlying capabilities are not stable: A person might encounter many chat interfaces with differing capabilities---for example, one provided by their organization with access to internal information stores but not the internet, or an agent running on their local machine that has access to their file system. Across these contexts, and as LLM-based tooling evolves, users must continuously revise their mental model of what a tool can do and, sometimes, how it does it. As long as capabilities keep changing, we expect how users understand them to remain an essential subject of study.

\section*{Data Availability and Ethics}
WildChat-1M is released by its creators under the ODC-BY license \cite{AllenAI2024WildChat1MCard}; users of the chatbot accepted terms permitting collection and release, and the dataset was de-identified by its creators, with IP addresses hashed \cite{Zhao2024WildChat:Wild}. We make no attempt to re-identify users, treat hashed IPs only as an anonymous proxy for recurrence, and redact personal names that survived de-identification in quoted prompts. Our replication package (\url{https://github.com/elleobrien/wildchat-misconceptions-replication}) contains the corpus filter, all prompts and run scripts, the conversation identifier map, and the final labels; transcripts are obtained from the WildChat release under its license.

\begin{acks}
We thank Mircea Lungu, who served as a validation coder in the preliminary study (Appendix~\ref{sec:preliminary-study}).
\end{acks}

\bibliographystyle{ACM-Reference-Format}
\bibliography{references}


\begin{thebibliography}{73}


\ifx \showCODEN    \undefined \def \showCODEN     #1{\unskip}     \fi
\ifx \showISBNx    \undefined \def \showISBNx     #1{\unskip}     \fi
\ifx \showISBNxiii \undefined \def \showISBNxiii  #1{\unskip}     \fi
\ifx \showISSN     \undefined \def \showISSN      #1{\unskip}     \fi
\ifx \showLCCN     \undefined \def \showLCCN      #1{\unskip}     \fi
\ifx \shownote     \undefined \def \shownote      #1{#1}          \fi
\ifx \showarticletitle \undefined \def \showarticletitle #1{#1}   \fi
\ifx \showURL      \undefined \def \showURL       {\relax}        \fi
\providecommand\bibfield[2]{#2}
\providecommand\bibinfo[2]{#2}
\providecommand\natexlab[1]{#1}
\providecommand\showeprint[2][]{arXiv:#2}

\bibitem[{Allen Institute for AI}(2024)]%
        {AllenAI2024WildChat1MCard}
\bibfield{author}{\bibinfo{person}{{Allen Institute for AI}}.}
  \bibinfo{year}{2024}\natexlab{}.
\newblock \bibinfo{title}{{allenai/WildChat-1M}}.
\newblock
  \bibinfo{howpublished}{\url{https://huggingface.co/datasets/allenai/WildChat-1M}}.
\newblock
\shownote{Hugging Face dataset card; license: ODC-BY; accessed 2026-10-05}.
\newblock


\bibitem[Amershi et~al\mbox{.}(2019)]%
        {Amershi2019GuidelinesInteraction}
\bibfield{author}{\bibinfo{person}{Saleema Amershi}, \bibinfo{person}{Dan
  Weld}, \bibinfo{person}{Mihaela Vorvoreanu}, \bibinfo{person}{Adam Fourney},
  \bibinfo{person}{Besmira Nushi}, \bibinfo{person}{Penny Collisson},
  \bibinfo{person}{Jina Suh}, \bibinfo{person}{Shamsi Iqbal},
  \bibinfo{person}{Paul~N. Bennett}, \bibinfo{person}{Kori Inkpen},
  \bibinfo{person}{Jaime Teevan}, \bibinfo{person}{Ruth Kikin-Gil}, {and}
  \bibinfo{person}{Eric Horvitz}.} \bibinfo{year}{2019}\natexlab{}.
\newblock \showarticletitle{{Guidelines for Human-AI Interaction}}. In
  \bibinfo{booktitle}{\emph{Proceedings of the 2019 CHI Conference on Human
  Factors in Computing Systems (CHI '19)}}. \bibinfo{publisher}{ACM}.
\newblock
\href{https://doi.org/10.1145/3290605.3300233}{doi:\nolinkurl{10.1145/3290605.3300233}}


\bibitem[Baltes et~al\mbox{.}(2025)]%
        {Baltes2025GuidelinesModels}
\bibfield{author}{\bibinfo{person}{Sebastian Baltes}, \bibinfo{person}{Florian
  Angermeir}, \bibinfo{person}{Chetan Arora}, \bibinfo{person}{Marvin
  {Mu\~{n}oz Bar\'{o}n}}, \bibinfo{person}{Chunyang Chen},
  \bibinfo{person}{Lukas B\"{o}hme}, \bibinfo{person}{Fabio Calefato},
  \bibinfo{person}{Neil Ernst}, \bibinfo{person}{Davide Falessi},
  \bibinfo{person}{Brian Fitzgerald}, \bibinfo{person}{Davide Fucci},
  \bibinfo{person}{Junda He}, \bibinfo{person}{Christoph Treude},
  \bibinfo{person}{Marcos Kalinowski}, \bibinfo{person}{Stefano Lambiase},
  \bibinfo{person}{Daniel Russo}, \bibinfo{person}{Mircea Lungu},
  \bibinfo{person}{Cristina {Martinez Montes}}, \bibinfo{person}{Lutz
  Prechelt}, \bibinfo{person}{Paul Ralph}, \bibinfo{person}{Rijnard {van
  Tonder}}, {and} \bibinfo{person}{Stefan Wagner}.}
  \bibinfo{year}{2025}\natexlab{}.
\newblock \bibinfo{title}{{Guidelines for Empirical Studies in Software
  Engineering involving Large Language Models}}.
\newblock \bibinfo{howpublished}{arXiv preprint arXiv:2508.15503}.
\newblock


\bibitem[Bansal et~al\mbox{.}(2019)]%
        {Bansal2019BeyondPerformance}
\bibfield{author}{\bibinfo{person}{Gagan Bansal}, \bibinfo{person}{Besmira
  Nushi}, \bibinfo{person}{Ece Kamar}, \bibinfo{person}{Walter~S. Lasecki},
  \bibinfo{person}{Daniel~S. Weld}, {and} \bibinfo{person}{Eric Horvitz}.}
  \bibinfo{year}{2019}\natexlab{}.
\newblock \showarticletitle{{Beyond Accuracy: The Role of Mental Models in
  Human-AI Team Performance}}.
\newblock \bibinfo{journal}{\emph{Proceedings of the AAAI Conference on Human
  Computation and Crowdsourcing}}  \bibinfo{volume}{7} (\bibinfo{year}{2019}),
  \bibinfo{pages}{2--11}.
\newblock
\showISBNx{9781577358206}
\showISSN{27691349}
\href{https://doi.org/10.1609/HCOMP.V7I1.5285}{doi:\nolinkurl{10.1609/HCOMP.V7I1.5285}}


\bibitem[Barke et~al\mbox{.}(2023)]%
        {Barke2023GroundedModels}
\bibfield{author}{\bibinfo{person}{Shraddha Barke}, \bibinfo{person}{Michael~B.
  James}, {and} \bibinfo{person}{Nadia Polikarpova}.}
  \bibinfo{year}{2023}\natexlab{}.
\newblock \showarticletitle{{Grounded Copilot: How Programmers Interact with
  Code-Generating Models}}.
\newblock \bibinfo{journal}{\emph{Proceedings of the ACM on Programming
  Languages}} \bibinfo{volume}{7}, \bibinfo{number}{OOPSLA1} (\bibinfo{date}{4}
  \bibinfo{year}{2023}).
\newblock
\showISSN{24751421}
\href{https://doi.org/10.1145/3586030}{doi:\nolinkurl{10.1145/3586030}}


\bibitem[Becker et~al\mbox{.}(2025)]%
        {BeckerMeasuringProductivity}
\bibfield{author}{\bibinfo{person}{Joel Becker}, \bibinfo{person}{Nate Rush},
  \bibinfo{person}{Beth Barnes}, {and} \bibinfo{person}{David Rein}.}
  \bibinfo{year}{2025}\natexlab{}.
\newblock \bibinfo{title}{{Measuring the Impact of Early-2025 AI on Experienced
  Open-Source Developer Productivity}}.
\newblock \bibinfo{howpublished}{arXiv preprint arXiv:2507.09089}.
\newblock


\bibitem[Brachman et~al\mbox{.}(2025)]%
        {Brachman2025BuildingChat-bot}
\bibfield{author}{\bibinfo{person}{Michelle Brachman}, \bibinfo{person}{Siya
  Kunde}, \bibinfo{person}{Sarah Miller}, \bibinfo{person}{Ana Fucs},
  \bibinfo{person}{Samantha Dempsey}, \bibinfo{person}{Jamie Jabbour}, {and}
  \bibinfo{person}{Werner Geyer}.} \bibinfo{year}{2025}\natexlab{}.
\newblock \showarticletitle{{Building Appropriate Mental Models: What Users
  Know and Want to Know about an Agentic AI Chatbot}}. In
  \bibinfo{booktitle}{\emph{Proceedings of the 30th International Conference on
  Intelligent User Interfaces}} \emph{(\bibinfo{series}{IUI '25})}.
  \bibinfo{publisher}{ACM}.
\newblock
\showISBNx{9798400713064}
\href{https://doi.org/10.1145/3708359.3712071}{doi:\nolinkurl{10.1145/3708359.3712071}}


\bibitem[Brachman et~al\mbox{.}(2023)]%
        {BrachmanFollowSystems}
\bibfield{author}{\bibinfo{person}{Michelle Brachman}, \bibinfo{person}{Qian
  Pan}, \bibinfo{person}{Hyo~Jin Do}, \bibinfo{person}{Casey Dugan},
  \bibinfo{person}{Arunima Chaudhary}, \bibinfo{person}{James~M. Johnson},
  \bibinfo{person}{Priyanshu Rai}, \bibinfo{person}{Thomas Gschwind},
  \bibinfo{person}{Jim Laredo}, \bibinfo{person}{Christoph Miksovic},
  \bibinfo{person}{Paolo Scotton}, \bibinfo{person}{Kartik Talamadupula},
  \bibinfo{person}{Gegi Thomas}, {and} \bibinfo{person}{Tathagata
  Chakraborti}.} \bibinfo{year}{2023}\natexlab{}.
\newblock \showarticletitle{{Follow the Successful Herd: Towards Explanations
  for Improved Use and Mental Models of Natural Language Systems}}. In
  \bibinfo{booktitle}{\emph{Proceedings of the 28th International Conference on
  Intelligent User Interfaces}} \emph{(\bibinfo{series}{IUI '23})}.
  \bibinfo{publisher}{ACM}.
\newblock
\showISBNx{9798400701061}
\href{https://doi.org/10.1145/3581641.3584088}{doi:\nolinkurl{10.1145/3581641.3584088}}


\bibitem[Champa et~al\mbox{.}(2024)]%
        {Champa2024ChatGPTDevelopment}
\bibfield{author}{\bibinfo{person}{Arifa~Islam Champa},
  \bibinfo{person}{Md~Fazle Rabbi}, \bibinfo{person}{Costain Nachuma}, {and}
  \bibinfo{person}{Minhaz~F. Zibran}.} \bibinfo{year}{2024}\natexlab{}.
\newblock \showarticletitle{{ChatGPT in Action: Analyzing Its Use in Software
  Development}}. In \bibinfo{booktitle}{\emph{Proceedings of the 21st
  International Conference on Mining Software Repositories}}
  \emph{(\bibinfo{series}{MSR '24})}. \bibinfo{publisher}{Association for
  Computing Machinery}, \bibinfo{pages}{182--186}.
\newblock
\href{https://doi.org/10.1145/3643991.3645077}{doi:\nolinkurl{10.1145/3643991.3645077}}


\bibitem[Chen et~al\mbox{.}(2021)]%
        {Chen2021EvaluatingCode}
\bibfield{author}{\bibinfo{person}{Mark Chen}, \bibinfo{person}{Jerry Tworek},
  \bibinfo{person}{Heewoo Jun}, \bibinfo{person}{Qiming Yuan},
  \bibinfo{person}{Henrique Ponde de~Oliveira Pinto}, \bibinfo{person}{Jared
  Kaplan}, \bibinfo{person}{Harri Edwards}, \bibinfo{person}{Yuri Burda},
  \bibinfo{person}{Nicholas Joseph}, \bibinfo{person}{Greg Brockman},
  \bibinfo{person}{Alex Ray}, \bibinfo{person}{Raul Puri},
  \bibinfo{person}{Gretchen Krueger}, \bibinfo{person}{Michael Petrov},
  \bibinfo{person}{Heidy Khlaaf}, \bibinfo{person}{Girish Sastry},
  \bibinfo{person}{Pamela Mishkin}, \bibinfo{person}{Brooke Chan},
  \bibinfo{person}{Scott Gray}, \bibinfo{person}{Nick Ryder},
  \bibinfo{person}{Mikhail Pavlov}, \bibinfo{person}{Alethea Power},
  \bibinfo{person}{Lukasz Kaiser}, \bibinfo{person}{Mohammad Bavarian},
  \bibinfo{person}{Clemens Winter}, \bibinfo{person}{Philippe Tillet},
  \bibinfo{person}{Felipe~Petroski Such}, \bibinfo{person}{Dave Cummings},
  \bibinfo{person}{Matthias Plappert}, \bibinfo{person}{Fotios Chantzis},
  \bibinfo{person}{Elizabeth Barnes}, \bibinfo{person}{Ariel Herbert-Voss},
  \bibinfo{person}{William~Hebgen Guss}, \bibinfo{person}{Alex Nichol},
  \bibinfo{person}{Alex Paino}, \bibinfo{person}{Nikolas Tezak},
  \bibinfo{person}{Jie Tang}, \bibinfo{person}{Igor Babuschkin},
  \bibinfo{person}{Suchir Balaji}, \bibinfo{person}{Shantanu Jain},
  \bibinfo{person}{William Saunders}, \bibinfo{person}{Christopher Hesse},
  \bibinfo{person}{Andrew~N. Carr}, \bibinfo{person}{Jan Leike},
  \bibinfo{person}{Josh Achiam}, \bibinfo{person}{Vedant Misra},
  \bibinfo{person}{Evan Morikawa}, \bibinfo{person}{Alec Radford},
  \bibinfo{person}{Matthew Knight}, \bibinfo{person}{Miles Brundage},
  \bibinfo{person}{Mira Murati}, \bibinfo{person}{Katie Mayer},
  \bibinfo{person}{Peter Welinder}, \bibinfo{person}{Bob McGrew},
  \bibinfo{person}{Dario Amodei}, \bibinfo{person}{Sam McCandlish},
  \bibinfo{person}{Ilya Sutskever}, {and} \bibinfo{person}{Wojciech Zaremba}.}
  \bibinfo{year}{2021}\natexlab{}.
\newblock \bibinfo{title}{{Evaluating Large Language Models Trained on Code}}.
\newblock \bibinfo{howpublished}{arXiv preprint arXiv:2107.03374}.
\newblock


\bibitem[Cheng et~al\mbox{.}(2024)]%
        {Cheng2024ItTools}
\bibfield{author}{\bibinfo{person}{Ruijia Cheng}, \bibinfo{person}{Ruotong
  Wang}, \bibinfo{person}{Thomas Zimmermann}, {and} \bibinfo{person}{Denae
  Ford}.} \bibinfo{year}{2024}\natexlab{}.
\newblock \showarticletitle{{``It would work for me too'': How online
  communities shape software developers' trust in AI-powered code generation
  tools}}.
\newblock \bibinfo{journal}{\emph{ACM Transactions on Interactive Intelligent
  Systems}} \bibinfo{volume}{14}, \bibinfo{number}{2} (\bibinfo{year}{2024}),
  \bibinfo{pages}{1--39}.
\newblock
\showISSN{21606463}
\href{https://doi.org/10.1145/3651990}{doi:\nolinkurl{10.1145/3651990}}


\bibitem[Chowdhury et~al\mbox{.}(2025)]%
        {Chowdhury2025InvestigatingAI}
\bibfield{author}{\bibinfo{person}{Neil Chowdhury}, \bibinfo{person}{Daniel
  Johnson}, \bibinfo{person}{Vincent Huang}, \bibinfo{person}{Jacob
  Steinhardt}, {and} \bibinfo{person}{Sarah Schwettmann}.}
  \bibinfo{year}{2025}\natexlab{}.
\newblock \bibinfo{title}{{Investigating Truthfulness in a Pre-Release o3
  Model}}.
\newblock \bibinfo{howpublished}{Transluce technical report}.
\newblock
\urldef\tempurl%
\url{https://transluce.org/investigating-o3-truthfulness}
\showURL{%
\tempurl}


\bibitem[Deng(2023)]%
        {Deng2023ChatGPTSpace}
\bibfield{author}{\bibinfo{person}{Yuntian Deng}.}
  \bibinfo{year}{2023}\natexlab{}.
\newblock \bibinfo{title}{{ChatGPT (Hugging Face Space)}}.
\newblock
  \bibinfo{howpublished}{\url{https://huggingface.co/spaces/yuntian-deng/ChatGPT}}.
\newblock
\shownote{Source code of the chatbot used to collect WildChat; accessed
  2026-09-08. The current revision targets a newer model; the API payload
  structure is unchanged}.
\newblock


\bibitem[Do et~al\mbox{.}(2024)]%
        {Do2024EvaluatingSystems}
\bibfield{author}{\bibinfo{person}{Hyo~Jin Do}, \bibinfo{person}{Michelle
  Brachman}, \bibinfo{person}{Casey Dugan}, \bibinfo{person}{Qian Pan},
  \bibinfo{person}{Priyanshu Rai}, \bibinfo{person}{James~M. Johnson}, {and}
  \bibinfo{person}{Roshni Thawani}.} \bibinfo{year}{2024}\natexlab{}.
\newblock \showarticletitle{{Evaluating What Others Say: The Effect of Accuracy
  Assessment in Shaping Mental Models of AI Systems}}.
\newblock \bibinfo{journal}{\emph{Proceedings of the ACM on Human-Computer
  Interaction}} \bibinfo{volume}{8}, \bibinfo{number}{CSCW2}
  (\bibinfo{date}{11} \bibinfo{year}{2024}), \bibinfo{pages}{373}.
\newblock
\showISSN{25730142}
\href{https://doi.org/10.1145/3686912}{doi:\nolinkurl{10.1145/3686912}}


\bibitem[Dunivin(2025)]%
        {Dunivin2025ScalingAnalysis}
\bibfield{author}{\bibinfo{person}{Zackary~Okun Dunivin}.}
  \bibinfo{year}{2025}\natexlab{}.
\newblock \showarticletitle{{Scaling hermeneutics: A guide to qualitative
  coding with LLMs for reflexive content analysis}}.
\newblock \bibinfo{journal}{\emph{EPJ Data Science}} \bibinfo{volume}{14},
  \bibinfo{number}{1} (\bibinfo{year}{2025}), \bibinfo{pages}{28}.
\newblock
\href{https://doi.org/10.1140/epjds/s13688-025-00548-8}{doi:\nolinkurl{10.1140/epjds/s13688-025-00548-8}}


\bibitem[Ernst and Treude(2026)]%
        {Ernst2026GenAIEngineering}
\bibfield{author}{\bibinfo{person}{Neil~A. Ernst} {and}
  \bibinfo{person}{Christoph Treude}.} \bibinfo{year}{2026}\natexlab{}.
\newblock \bibinfo{title}{{GenAI is no silver bullet for qualitative research
  in software engineering}}.
\newblock \bibinfo{howpublished}{arXiv preprint arXiv:2603.08951}.
\newblock
\href{https://doi.org/10.48550/arXiv.2603.08951}{doi:\nolinkurl{10.48550/arXiv.2603.08951}}


\bibitem[Fawzy et~al\mbox{.}(2025)]%
        {Fawzy2025VibeReview}
\bibfield{author}{\bibinfo{person}{Ahmed Fawzy}, \bibinfo{person}{Amjed Tahir},
  {and} \bibinfo{person}{Kelly Blincoe}.} \bibinfo{year}{2025}\natexlab{}.
\newblock \bibinfo{title}{{Vibe Coding in Practice: Motivations, Challenges,
  and a Future Outlook - a Grey Literature Review}}.
\newblock \bibinfo{howpublished}{arXiv preprint arXiv:2510.00328}.
\newblock


\bibitem[Feldman and Anderson(2024)]%
        {Feldman2024Non-ExpertFuture}
\bibfield{author}{\bibinfo{person}{Molly~Q. Feldman} {and}
  \bibinfo{person}{Carolyn~Jane Anderson}.} \bibinfo{year}{2024}\natexlab{}.
\newblock \showarticletitle{{Non-Expert Programmers in the Generative AI
  Future}}.
\newblock \bibinfo{journal}{\emph{ACM International Conference Proceeding
  Series}} (\bibinfo{date}{6} \bibinfo{year}{2024}).
\newblock
\showISBNx{9798400710179}
\href{https://doi.org/10.1145/3663384.3663393}{doi:\nolinkurl{10.1145/3663384.3663393}}


\bibitem[Ferdowsifard et~al\mbox{.}(2020)]%
        {Ferdowsifard2020Small-stepExample}
\bibfield{author}{\bibinfo{person}{Kasra Ferdowsifard}, \bibinfo{person}{Allen
  Ordookhanians}, \bibinfo{person}{Hila Peleg}, \bibinfo{person}{Sorin Lerner},
  {and} \bibinfo{person}{Nadia Polikarpova}.} \bibinfo{year}{2020}\natexlab{}.
\newblock \showarticletitle{{Small-step live programming by example}}. In
  \bibinfo{booktitle}{\emph{Proceedings of the 33rd Annual ACM Symposium on
  User Interface Software and Technology (UIST '20)}}.
  \bibinfo{publisher}{Association for Computing Machinery, Inc},
  \bibinfo{pages}{614--626}.
\newblock
\showISBNx{9781450375146}
\href{https://doi.org/10.1145/3379337.3415869}{doi:\nolinkurl{10.1145/3379337.3415869}}


\bibitem[Gadala(2017)]%
        {Gadala2017AutomationStrategies}
\bibfield{author}{\bibinfo{person}{Marwa Gadala}.}
  \bibinfo{year}{2017}\natexlab{}.
\newblock \emph{\bibinfo{title}{{Automation bias: exploring causal mechanisms
  and potential mitigation strategies}}}.
\newblock Doctoral thesis. \bibinfo{school}{City, University of London}.
\newblock
\urldef\tempurl%
\url{https://openaccess.city.ac.uk/id/eprint/17889/}
\showURL{%
\tempurl}


\bibitem[Gero et~al\mbox{.}(2020)]%
        {Gero2020MentalSetting}
\bibfield{author}{\bibinfo{person}{Katy~Ilonka Gero}, \bibinfo{person}{Zahra
  Ashktorab}, \bibinfo{person}{Casey Dugan}, \bibinfo{person}{Qian Pan},
  \bibinfo{person}{James Johnson}, \bibinfo{person}{Werner Geyer},
  \bibinfo{person}{Maria Ruiz}, \bibinfo{person}{Sarah Miller},
  \bibinfo{person}{David~R. Millen}, \bibinfo{person}{Murray Campbell},
  \bibinfo{person}{Sadhana Kumaravel}, {and} \bibinfo{person}{Wei Zhang}.}
  \bibinfo{year}{2020}\natexlab{}.
\newblock \showarticletitle{{Mental Models of AI Agents in a Cooperative Game
  Setting}}. In \bibinfo{booktitle}{\emph{Proceedings of the 2020 CHI
  Conference on Human Factors in Computing Systems (CHI '20)}}.
  \bibinfo{publisher}{Association for Computing Machinery}.
\newblock
\showISBNx{9781450367080}
\href{https://doi.org/10.1145/3313831.3376316}{doi:\nolinkurl{10.1145/3313831.3376316}}


\bibitem[Goddard et~al\mbox{.}(2012)]%
        {Goddard2012AutomationMitigators}
\bibfield{author}{\bibinfo{person}{Kate Goddard}, \bibinfo{person}{Abdul
  Roudsari}, {and} \bibinfo{person}{Jeremy~C. Wyatt}.}
  \bibinfo{year}{2012}\natexlab{}.
\newblock \showarticletitle{{Automation bias: A systematic review of frequency,
  effect mediators, and mitigators}}.
\newblock \bibinfo{journal}{\emph{Journal of the American Medical Informatics
  Association}} \bibinfo{volume}{19}, \bibinfo{number}{1} (\bibinfo{date}{1}
  \bibinfo{year}{2012}), \bibinfo{pages}{121--127}.
\newblock
\showISSN{10675027}
\href{https://doi.org/10.1136/amiajnl-2011-000089}{doi:\nolinkurl{10.1136/amiajnl-2011-000089}}


\bibitem[Hicke and Tomlinson(2026)]%
        {Hicke2026AdoptWild}
\bibfield{author}{\bibinfo{person}{Rebecca M.~M. Hicke} {and}
  \bibinfo{person}{Kiran Tomlinson}.} \bibinfo{year}{2026}\natexlab{}.
\newblock \bibinfo{title}{{Adopt ≠ Adapt: Longitudinal Analyses of LLM
  Conversations in the Wild}}.
\newblock \bibinfo{howpublished}{arXiv preprint arXiv:2605.29018}.
\newblock


\bibitem[Jayagopal et~al\mbox{.}(2022)]%
        {Jayagopal2022ExploringProgrammers}
\bibfield{author}{\bibinfo{person}{Dhanya Jayagopal}, \bibinfo{person}{Justin
  Lubin}, {and} \bibinfo{person}{Sarah~E. Chasins}.}
  \bibinfo{year}{2022}\natexlab{}.
\newblock \showarticletitle{{Exploring the Learnability of Program Synthesizers
  by Novice Programmers}}. In \bibinfo{booktitle}{\emph{Proceedings of the 35th
  Annual ACM Symposium on User Interface Software and Technology (UIST '22)}}.
  \bibinfo{publisher}{Association for Computing Machinery, Inc}.
\newblock
\showISBNx{9781450393201}
\href{https://doi.org/10.1145/3526113.3545659}{doi:\nolinkurl{10.1145/3526113.3545659}}


\bibitem[Kabir et~al\mbox{.}(2024)]%
        {Kabir2024IsQuestions}
\bibfield{author}{\bibinfo{person}{Samia Kabir}, \bibinfo{person}{David~N.
  Udo-Imeh}, \bibinfo{person}{Bonan Kou}, {and} \bibinfo{person}{Tianyi
  Zhang}.} \bibinfo{year}{2024}\natexlab{}.
\newblock \showarticletitle{{Is Stack Overflow Obsolete? An Empirical Study of
  the Characteristics of ChatGPT Answers to Stack Overflow Questions}}. In
  \bibinfo{booktitle}{\emph{Conference on Human Factors in Computing Systems}}.
  \bibinfo{publisher}{ACM}.
\newblock
\showISBNx{9798400703300}
\href{https://doi.org/10.1145/3613904.3642596}{doi:\nolinkurl{10.1145/3613904.3642596}}


\bibitem[Karpathy(2025)]%
        {Karpathy2025TheresCoding.}
\bibfield{author}{\bibinfo{person}{Andrej Karpathy}.}
  \bibinfo{year}{2025}\natexlab{}.
\newblock \bibinfo{title}{{There's a new kind of coding I call ``vibe
  coding''}}.
\newblock \bibinfo{howpublished}{Post on X (formerly Twitter), 2 February
  2025}.
\newblock
\shownote{Accessed 2026-09-15}.
\newblock
\urldef\tempurl%
\url{https://x.com/karpathy/status/1886192184808149383}
\showURL{%
\tempurl}


\bibitem[Kazemitabaar et~al\mbox{.}(2024)]%
        {Kazemitabaar2024CodeAid:Needs}
\bibfield{author}{\bibinfo{person}{Majeed Kazemitabaar},
  \bibinfo{person}{Runlong Ye}, \bibinfo{person}{Xiaoning Wang},
  \bibinfo{person}{Austin~Z. Henley}, \bibinfo{person}{Paul Denny},
  \bibinfo{person}{Michelle Craig}, {and} \bibinfo{person}{Tovi Grossman}.}
  \bibinfo{year}{2024}\natexlab{}.
\newblock \showarticletitle{{CodeAid: Evaluating a Classroom Deployment of an
  LLM-based Programming Assistant that Balances Student and Educator Needs}}.
  In \bibinfo{booktitle}{\emph{Proceedings of the 2024 CHI Conference on Human
  Factors in Computing Systems (CHI '24)}}. \bibinfo{publisher}{Association for
  Computing Machinery}.
\newblock
\showISBNx{9798400703300}
\href{https://doi.org/10.1145/3613904.3642773}{doi:\nolinkurl{10.1145/3613904.3642773}}


\bibitem[Kelly et~al\mbox{.}(2023)]%
        {Kelly2023CapturingApproach}
\bibfield{author}{\bibinfo{person}{Markelle Kelly}, \bibinfo{person}{Aakriti
  Kumar}, \bibinfo{person}{Padhraic Smyth}, {and} \bibinfo{person}{Mark
  Steyvers}.} \bibinfo{year}{2023}\natexlab{}.
\newblock \showarticletitle{{Capturing Humans' Mental Models of AI: An Item
  Response Theory Approach}}. In \bibinfo{booktitle}{\emph{Proceedings of the
  2023 ACM Conference on Fairness, Accountability, and Transparency (FAccT
  '23)}}. \bibinfo{publisher}{ACM}, \bibinfo{pages}{1723--1734}.
\newblock
\showISBNx{9781450372527}
\href{https://doi.org/10.1145/3593013.3594111}{doi:\nolinkurl{10.1145/3593013.3594111}}


\bibitem[Ko et~al\mbox{.}(2011)]%
        {KoTheEngineering}
\bibfield{author}{\bibinfo{person}{Andrew~J. Ko}, \bibinfo{person}{Robin
  Abraham}, \bibinfo{person}{Laura Beckwith}, \bibinfo{person}{Alan Blackwell},
  \bibinfo{person}{Margaret Burnett}, \bibinfo{person}{Martin Erwig},
  \bibinfo{person}{Chris Scaffidi}, \bibinfo{person}{Joseph Lawrance},
  \bibinfo{person}{Henry Lieberman}, \bibinfo{person}{Brad Myers},
  \bibinfo{person}{Mary~Beth Rosson}, \bibinfo{person}{Gregg Rothermel},
  \bibinfo{person}{Mary Shaw}, {and} \bibinfo{person}{Susan Wiedenbeck}.}
  \bibinfo{year}{2011}\natexlab{}.
\newblock \showarticletitle{{The State of the Art in End-User Software
  Engineering}}.
\newblock \bibinfo{journal}{\emph{Comput. Surveys}} \bibinfo{volume}{43},
  \bibinfo{number}{3} (\bibinfo{year}{2011}).
\newblock
\href{https://doi.org/10.1145/1922649.1922658}{doi:\nolinkurl{10.1145/1922649.1922658}}


\bibitem[Kulesza et~al\mbox{.}(2012)]%
        {Kulesza2012TellAgent}
\bibfield{author}{\bibinfo{person}{Todd Kulesza}, \bibinfo{person}{Simone
  Stumpf}, \bibinfo{person}{Margaret Burnett}, {and} \bibinfo{person}{Irwin
  Kwan}.} \bibinfo{year}{2012}\natexlab{}.
\newblock \showarticletitle{{Tell me more? the effects of mental model
  soundness on personalizing an intelligent agent}}. In
  \bibinfo{booktitle}{\emph{Proceedings of the SIGCHI Conference on Human
  Factors in Computing Systems (CHI '12)}}. \bibinfo{pages}{1--10}.
\newblock
\showISBNx{9781450310154}
\href{https://doi.org/10.1145/2207676.2207678}{doi:\nolinkurl{10.1145/2207676.2207678}}


\bibitem[Landis and Koch(1977)]%
        {Landis1977TheData}
\bibfield{author}{\bibinfo{person}{J.~Richard Landis} {and}
  \bibinfo{person}{Gary~G. Koch}.} \bibinfo{year}{1977}\natexlab{}.
\newblock \showarticletitle{{The Measurement of Observer Agreement for
  Categorical Data}}.
\newblock \bibinfo{journal}{\emph{Biometrics}} \bibinfo{volume}{33},
  \bibinfo{number}{1} (\bibinfo{date}{3} \bibinfo{year}{1977}),
  \bibinfo{pages}{159--174}.
\newblock
\showISSN{0006341X}
\href{https://doi.org/10.2307/2529310}{doi:\nolinkurl{10.2307/2529310}}


\bibitem[Lau and Guo(2023)]%
        {Lau2023FromCopilot}
\bibfield{author}{\bibinfo{person}{Sam Lau} {and} \bibinfo{person}{Philip
  Guo}.} \bibinfo{year}{2023}\natexlab{}.
\newblock \showarticletitle{{From "Ban It Till We Understand It" to "Resistance
  is Futile": How University Programming Instructors Plan to Adapt as More
  Students Use AI Code Generation and Explanation Tools such as ChatGPT and
  GitHub Copilot}}. In \bibinfo{booktitle}{\emph{Proceedings of the 2023 ACM
  Conference on International Computing Education Research - Volume 1 (ICER
  '23)}}. \bibinfo{publisher}{Association for Computing Machinery, Inc},
  \bibinfo{pages}{106--121}.
\newblock
\showISBNx{9781450399760}
\href{https://doi.org/10.1145/3568813.3600138}{doi:\nolinkurl{10.1145/3568813.3600138}}


\bibitem[Lee et~al\mbox{.}(2025)]%
        {Lee2025TheWorkers}
\bibfield{author}{\bibinfo{person}{H.~P.~H. Lee}, \bibinfo{person}{A. Sarkar},
  \bibinfo{person}{L. Tankelevitch}, \bibinfo{person}{I. Drosos},
  \bibinfo{person}{S. Rintel}, \bibinfo{person}{R. Banks}, {and}
  \bibinfo{person}{N. Wilson}.} \bibinfo{year}{2025}\natexlab{}.
\newblock \showarticletitle{{The Impact of Generative AI on Critical Thinking:
  Self-Reported Reductions in Cognitive Effort and Confidence Effects From a
  Survey of Knowledge Workers}}. In \bibinfo{booktitle}{\emph{Proceedings of
  the 2025 CHI Conference on Human Factors in Computing Systems (CHI '25)}}.
  \bibinfo{publisher}{ACM}, \bibinfo{pages}{1--22}.
\newblock
\href{https://doi.org/10.1145/3706598.3713778}{doi:\nolinkurl{10.1145/3706598.3713778}}


\bibitem[Lehtinen et~al\mbox{.}(2024)]%
        {Lehtinen2024LetsQuestions}
\bibfield{author}{\bibinfo{person}{Teemu Lehtinen}, \bibinfo{person}{Charles
  Koutcheme}, {and} \bibinfo{person}{Arto Hellas}.}
  \bibinfo{year}{2024}\natexlab{}.
\newblock \showarticletitle{{Let’s Ask AI About Their Programs: Exploring
  ChatGPT’s Answers To Program Comprehension Questions}}. In
  \bibinfo{booktitle}{\emph{Proceedings of the 46th International Conference on
  Software Engineering: Software Engineering Education and Training (ICSE-SEET
  '24)}}. \bibinfo{publisher}{IEEE Computer Society},
  \bibinfo{pages}{221--232}.
\newblock
\showISBNx{9798400704987}
\showISSN{02705257}
\href{https://doi.org/10.1145/3639474.3640058}{doi:\nolinkurl{10.1145/3639474.3640058}}


\bibitem[Liang et~al\mbox{.}(2024)]%
        {Liang2024AChallenges}
\bibfield{author}{\bibinfo{person}{Jenny~T. Liang}, \bibinfo{person}{Chenyang
  Yang}, {and} \bibinfo{person}{Brad~A. Myers}.}
  \bibinfo{year}{2024}\natexlab{}.
\newblock \showarticletitle{{A Large-Scale Survey on the Usability of AI
  Programming Assistants: Successes and Challenges}}. In
  \bibinfo{booktitle}{\emph{Proceedings of the 46th IEEE/ACM International
  Conference on Software Engineering (ICSE '24)}}. \bibinfo{publisher}{IEEE
  Computer Society}.
\newblock
\showISBNx{9798400702174}
\showISSN{02705257}
\href{https://doi.org/10.1145/3597503.3608128}{doi:\nolinkurl{10.1145/3597503.3608128}}


\bibitem[Liu et~al\mbox{.}(2023)]%
        {Liu2023WhatItWants}
\bibfield{author}{\bibinfo{person}{Michael~Xieyang Liu},
  \bibinfo{person}{Advait Sarkar}, \bibinfo{person}{Carina Negreanu},
  \bibinfo{person}{Benjamin Zorn}, \bibinfo{person}{Jack Williams},
  \bibinfo{person}{Neil Toronto}, {and} \bibinfo{person}{Andrew~D. Gordon}.}
  \bibinfo{year}{2023}\natexlab{}.
\newblock \showarticletitle{{``What It Wants Me To Say'': Bridging the
  Abstraction Gap Between End-User Programmers and Code-Generating Large
  Language Models}}. In \bibinfo{booktitle}{\emph{Proceedings of the 2023 CHI
  Conference on Human Factors in Computing Systems}}.
  \bibinfo{publisher}{Association for Computing Machinery},
  \bibinfo{numpages}{31}~pages.
\newblock
\showISBNx{9781450394215}
\href{https://doi.org/10.1145/3544548.3580817}{doi:\nolinkurl{10.1145/3544548.3580817}}


\bibitem[Liu et~al\mbox{.}(2024)]%
        {Liu2024LostContexts}
\bibfield{author}{\bibinfo{person}{Nelson~F. Liu}, \bibinfo{person}{Kevin Lin},
  \bibinfo{person}{John Hewitt}, \bibinfo{person}{Ashwin Paranjape},
  \bibinfo{person}{Michele Bevilacqua}, \bibinfo{person}{Fabio Petroni}, {and}
  \bibinfo{person}{Percy Liang}.} \bibinfo{year}{2024}\natexlab{}.
\newblock \showarticletitle{{Lost in the Middle: How Language Models Use Long
  Contexts}}.
\newblock \bibinfo{journal}{\emph{Transactions of the Association for
  Computational Linguistics}}  \bibinfo{volume}{12} (\bibinfo{year}{2024}),
  \bibinfo{pages}{157--173}.
\newblock
\showISSN{2307387X}
\href{https://doi.org/10.1162/TACL{\_}A{\_}00638}{doi:\nolinkurl{10.1162/TACL{\_}A{\_}00638}}


\bibitem[McCandless et~al\mbox{.}(2024)]%
        {McCandless2024LiberatingStructures}
\bibfield{author}{\bibinfo{person}{Keith McCandless}, \bibinfo{person}{Arvind
  Singhal}, {and} \bibinfo{person}{Steven~H. Cady}.}
  \bibinfo{year}{2024}\natexlab{}.
\newblock \showarticletitle{{Liberating structures}}.
\newblock In \bibinfo{booktitle}{\emph{Elgar Encyclopedia of
  Interdisciplinarity and Transdisciplinarity}}. \bibinfo{publisher}{Edward
  Elgar Publishing}, \bibinfo{pages}{318--322}.
\newblock
\href{https://doi.org/10.4337/9781035317967.ch70}{doi:\nolinkurl{10.4337/9781035317967.ch70}}


\bibitem[Mirzadeh et~al\mbox{.}(2024)]%
        {Mirzadeh2024GSM-Symbolic:Models}
\bibfield{author}{\bibinfo{person}{Iman Mirzadeh}, \bibinfo{person}{Keivan
  Alizadeh}, \bibinfo{person}{Hooman Shahrokhi}, \bibinfo{person}{Oncel Tuzel},
  \bibinfo{person}{Samy Bengio}, {and} \bibinfo{person}{Mehrdad Farajtabar}.}
  \bibinfo{year}{2024}\natexlab{}.
\newblock \bibinfo{title}{{GSM-Symbolic: Understanding the Limitations of
  Mathematical Reasoning in Large Language Models}}.
\newblock \bibinfo{howpublished}{arXiv preprint arXiv:2410.05229}.
\newblock


\bibitem[Mohamed et~al\mbox{.}(2024)]%
        {Mohamed2024ChattingChatGPT}
\bibfield{author}{\bibinfo{person}{Suad Mohamed}, \bibinfo{person}{Abdullah
  Parvin}, {and} \bibinfo{person}{Esteban Parra}.}
  \bibinfo{year}{2024}\natexlab{}.
\newblock \showarticletitle{{Chatting with AI: Deciphering Developer
  Conversations with ChatGPT}}. In \bibinfo{booktitle}{\emph{Proceedings of the
  21st International Conference on Mining Software Repositories}}
  \emph{(\bibinfo{series}{MSR '24})}. \bibinfo{publisher}{Association for
  Computing Machinery}, \bibinfo{pages}{187--191}.
\newblock
\href{https://doi.org/10.1145/3643991.3645078}{doi:\nolinkurl{10.1145/3643991.3645078}}


\bibitem[Nguyen et~al\mbox{.}(2024)]%
        {Nguyen2024HowOther}
\bibfield{author}{\bibinfo{person}{Sydney Nguyen},
  \bibinfo{person}{Hannah~McLean Babe}, \bibinfo{person}{Yangtian Zi},
  \bibinfo{person}{Arjun Guha}, \bibinfo{person}{Carolyn~Jane Anderson}, {and}
  \bibinfo{person}{Molly~Q Feldman}.} \bibinfo{year}{2024}\natexlab{}.
\newblock \showarticletitle{{How Beginning Programmers and Code LLMs (Mis) read
  Each Other}}. In \bibinfo{booktitle}{\emph{Proceedings of the CHI Conference
  on Human Factors in Computing Systems}} \emph{(\bibinfo{series}{CHI '24})}.
  \bibinfo{publisher}{Association for Computing Machinery},
  \bibinfo{pages}{1--26}.
\newblock
\href{https://doi.org/10.1145/3613904.3642706}{doi:\nolinkurl{10.1145/3613904.3642706}}


\bibitem[Norman(1988)]%
        {Norman1988TheThings}
\bibfield{author}{\bibinfo{person}{Donald~A. Norman}.}
  \bibinfo{year}{1988}\natexlab{}.
\newblock \bibinfo{booktitle}{\emph{{The Psychology of Everyday Things}}}.
\newblock \bibinfo{publisher}{Basic Books}.
\newblock
\showISBNx{978-0-465-06710-7}


\bibitem[Norman(1999)]%
        {Norman1999AffordanceDesign}
\bibfield{author}{\bibinfo{person}{Donald~A. Norman}.}
  \bibinfo{year}{1999}\natexlab{}.
\newblock \showarticletitle{{Affordance, conventions, and design}}.
\newblock \bibinfo{journal}{\emph{Interactions}} \bibinfo{volume}{6},
  \bibinfo{number}{3} (\bibinfo{date}{5} \bibinfo{year}{1999}),
  \bibinfo{pages}{38--43}.
\newblock


\bibitem[O'Brien(2025)]%
        {OBrien2025HowProgram}
\bibfield{author}{\bibinfo{person}{Gabrielle O'Brien}.}
  \bibinfo{year}{2025}\natexlab{}.
\newblock \showarticletitle{{How Scientists Use Large Language Models to
  Program}}. In \bibinfo{booktitle}{\emph{Proceedings of the 2025 CHI
  Conference on Human Factors in Computing Systems (CHI '25)}}.
  \bibinfo{publisher}{Association for Computing Machinery}.
\newblock
\href{https://doi.org/10.1145/3706598.3713668}{doi:\nolinkurl{10.1145/3706598.3713668}}


\bibitem[O'Brien et~al\mbox{.}(2026)]%
        {OBrien2026UserAssistants}
\bibfield{author}{\bibinfo{person}{Gabrielle O'Brien},
  \bibinfo{person}{Antonio~Pedro Santos~Alves}, \bibinfo{person}{Sebastian
  Baltes}, \bibinfo{person}{Grischa Liebel}, \bibinfo{person}{Mircea Lungu},
  {and} \bibinfo{person}{Marcos Kalinowski}.} \bibinfo{year}{2026}\natexlab{}.
\newblock \showarticletitle{{User Misconceptions of LLM-Based Conversational
  Programming Assistants}}. In \bibinfo{booktitle}{\emph{Proceedings of the
  ICSE 2026 Journal Ahead Workshop (JAWs)}}.
\newblock
\shownote{To appear}.
\newblock


\bibitem[Oliveira et~al\mbox{.}(2023)]%
        {Oliveira2023StudentMisconceptions}
\bibfield{author}{\bibinfo{person}{Eduardo Oliveira}, \bibinfo{person}{Hieke
  Keuning}, {and} \bibinfo{person}{Johan Jeuring}.}
  \bibinfo{year}{2023}\natexlab{}.
\newblock \showarticletitle{{Student Code Refactoring Misconceptions}}. In
  \bibinfo{booktitle}{\emph{Proceedings of the 2023 Conference on Innovation
  and Technology in Computer Science Education V. 1 (ITiCSE '23)}}.
  \bibinfo{publisher}{Association for Computing Machinery},
  \bibinfo{pages}{19--25}.
\newblock
\showISBNx{9798400701382}
\showISSN{1942647X}
\href{https://doi.org/10.1145/3587102.3588840}{doi:\nolinkurl{10.1145/3587102.3588840}}


\bibitem[{OpenAI}(2023)]%
        {OpenAI2023CodeInterpreter}
\bibfield{author}{\bibinfo{person}{{OpenAI}}.} \bibinfo{year}{2023}\natexlab{}.
\newblock \bibinfo{title}{{ChatGPT --- Release Notes}}.
\newblock
  \bibinfo{howpublished}{\url{https://help.openai.com/en/articles/6825453-chatgpt-release-notes}}.
\newblock
\shownote{Code interpreter beta for Plus users, July 2023; accessed
  2026-09-08}.
\newblock


\bibitem[Pimenova et~al\mbox{.}(2025)]%
        {Pimenova2025GoodCoding}
\bibfield{author}{\bibinfo{person}{Veronica Pimenova}, \bibinfo{person}{Sarah
  Fakhoury}, \bibinfo{person}{Christian Bird}, \bibinfo{person}{Margaret-Anne
  Storey}, {and} \bibinfo{person}{Madeline Endres}.}
  \bibinfo{year}{2025}\natexlab{}.
\newblock \bibinfo{title}{{Good Vibrations? A Qualitative Study of Co-Creation,
  Communication, Flow, and Trust in Vibe Coding}}.
\newblock \bibinfo{howpublished}{arXiv preprint arXiv:2509.12491}.
\newblock


\bibitem[Prather et~al\mbox{.}(2023)]%
        {Prather2023ItsProgrammers}
\bibfield{author}{\bibinfo{person}{James Prather}, \bibinfo{person}{Brent~N.
  Reeves}, \bibinfo{person}{Paul Denny}, \bibinfo{person}{Brett~A. Becker},
  \bibinfo{person}{Juho Leinonen}, \bibinfo{person}{Andrew Luxton-Reilly},
  \bibinfo{person}{Garrett Powell}, \bibinfo{person}{James Finnie-Ansley},
  {and} \bibinfo{person}{Eddie~Antonio Santos}.}
  \bibinfo{year}{2023}\natexlab{}.
\newblock \showarticletitle{{“It’s Weird That it Knows What I Want”:
  Usability and Interactions with Copilot for Novice Programmers}}.
\newblock \bibinfo{journal}{\emph{ACM Transactions on Computer-Human
  Interaction}} \bibinfo{volume}{31}, \bibinfo{number}{1} (\bibinfo{date}{11}
  \bibinfo{year}{2023}).
\newblock
\showISSN{15577325}
\href{https://doi.org/10.1145/3617367}{doi:\nolinkurl{10.1145/3617367}}


\bibitem[Prather et~al\mbox{.}(2024)]%
        {Prather2024TheProgrammers}
\bibfield{author}{\bibinfo{person}{James Prather}, \bibinfo{person}{Brent~N
  Reeves}, \bibinfo{person}{Juho Leinonen}, \bibinfo{person}{Stephen MacNeil},
  \bibinfo{person}{Arisoa~S Randrianasolo}, \bibinfo{person}{Brett~A. Becker},
  \bibinfo{person}{Bailey Kimmel}, \bibinfo{person}{Jared Wright}, {and}
  \bibinfo{person}{Ben Briggs}.} \bibinfo{year}{2024}\natexlab{}.
\newblock \showarticletitle{{The Widening Gap: The Benefits and Harms of
  Generative AI for Novice Programmers}}. In
  \bibinfo{booktitle}{\emph{Proceedings of the 2024 ACM Conference on
  International Computing Education Research - Volume 1 (ICER '24)}}.
  \bibinfo{publisher}{ACM}, \bibinfo{address}{New York, NY, USA},
  \bibinfo{pages}{469--486}.
\newblock
\showISBNx{9798400704758}
\href{https://doi.org/10.1145/3632620.3671116}{doi:\nolinkurl{10.1145/3632620.3671116}}


\bibitem[Qian and Lehman(2017)]%
        {Qian2017StudentsProgramming}
\bibfield{author}{\bibinfo{person}{Yizhou Qian} {and} \bibinfo{person}{James
  Lehman}.} \bibinfo{year}{2017}\natexlab{}.
\newblock \showarticletitle{{Students’ Misconceptions and Other Difficulties
  in Introductory Programming}}.
\newblock \bibinfo{journal}{\emph{ACM Transactions on Computing Education
  (TOCE)}} \bibinfo{volume}{18}, \bibinfo{number}{1} (\bibinfo{date}{10}
  \bibinfo{year}{2017}).
\newblock
\showISSN{19466226}
\href{https://doi.org/10.1145/3077618}{doi:\nolinkurl{10.1145/3077618}}


\bibitem[Sarkar et~al\mbox{.}(2022)]%
        {Sarkar2022WhatIsIt}
\bibfield{author}{\bibinfo{person}{Advait Sarkar}, \bibinfo{person}{Andrew~D.
  Gordon}, \bibinfo{person}{Carina Negreanu}, \bibinfo{person}{Christian
  Poelitz}, \bibinfo{person}{Sruti~Srinivasa Ragavan}, {and}
  \bibinfo{person}{Ben Zorn}.} \bibinfo{year}{2022}\natexlab{}.
\newblock \showarticletitle{{What is it like to program with artificial
  intelligence?}}. In \bibinfo{booktitle}{\emph{Proceedings of the 33rd Annual
  Workshop of the Psychology of Programming Interest Group (PPIG 2022)}}.
\newblock
\showeprint[arxiv]{2208.06213}
\urldef\tempurl%
\url{https://arxiv.org/abs/2208.06213}
\showURL{%
\tempurl}


\bibitem[Sarsa et~al\mbox{.}(2022)]%
        {Sarsa2022AutomaticModels}
\bibfield{author}{\bibinfo{person}{Sami Sarsa}, \bibinfo{person}{Paul Denny},
  \bibinfo{person}{Arto Hellas}, {and} \bibinfo{person}{Juho Leinonen}.}
  \bibinfo{year}{2022}\natexlab{}.
\newblock \showarticletitle{{Automatic Generation of Programming Exercises and
  Code Explanations Using Large Language Models}}. In
  \bibinfo{booktitle}{\emph{Proceedings of the 2022 ACM Conference on
  International Computing Education Research - Volume 1 (ICER '22)}}.
  \bibinfo{publisher}{Association for Computing Machinery, Inc},
  \bibinfo{pages}{27--43}.
\newblock
\showISBNx{9781450391948}
\href{https://doi.org/10.1145/3501385.3543957}{doi:\nolinkurl{10.1145/3501385.3543957}}


\bibitem[Shankar et~al\mbox{.}(2024)]%
        {Shankar2024WhoPreferences}
\bibfield{author}{\bibinfo{person}{Shreya Shankar}, \bibinfo{person}{J.~D.
  Zamfirescu-Pereira}, \bibinfo{person}{Bj\"{o}rn Hartmann},
  \bibinfo{person}{Aditya~G. Parameswaran}, {and} \bibinfo{person}{Ian
  Arawjo}.} \bibinfo{year}{2024}\natexlab{}.
\newblock \showarticletitle{{Who Validates the Validators? Aligning
  LLM-Assisted Evaluation of LLM Outputs with Human Preferences}}. In
  \bibinfo{booktitle}{\emph{UIST 2024 - Proceedings of the 37th Annual ACM
  Symposium on User Interface Software and Technology}}.
\newblock
\showISBNx{9798400706288}
\href{https://doi.org/10.1145/3654777.3676450}{doi:\nolinkurl{10.1145/3654777.3676450}}


\bibitem[Sharma et~al\mbox{.}(2024)]%
        {Sharma2023TowardsModels}
\bibfield{author}{\bibinfo{person}{Mrinank Sharma}, \bibinfo{person}{Meg Tong},
  \bibinfo{person}{Tomasz Korbak}, \bibinfo{person}{David Duvenaud},
  \bibinfo{person}{Amanda Askell}, \bibinfo{person}{Samuel~R. Bowman},
  \bibinfo{person}{Newton Cheng}, \bibinfo{person}{Esin Durmus},
  \bibinfo{person}{Zac Hatfield-Dodds}, \bibinfo{person}{Scott~R. Johnston},
  \bibinfo{person}{Shauna Kravec}, \bibinfo{person}{Timothy Maxwell},
  \bibinfo{person}{Sam McCandlish}, \bibinfo{person}{Kamal Ndousse},
  \bibinfo{person}{Oliver Rausch}, \bibinfo{person}{Nicholas Schiefer},
  \bibinfo{person}{Da Yan}, \bibinfo{person}{Miranda Zhang}, {and}
  \bibinfo{person}{Ethan Perez}.} \bibinfo{year}{2024}\natexlab{}.
\newblock \showarticletitle{{Towards Understanding Sycophancy in Language
  Models}}. In \bibinfo{booktitle}{\emph{The Twelfth International Conference
  on Learning Representations (ICLR)}}.
\newblock
\urldef\tempurl%
\url{https://arxiv.org/abs/2310.13548}
\showURL{%
\tempurl}


\bibitem[Smith~III et~al\mbox{.}(1994)]%
        {Iii1994MisconceptionsTransition}
\bibfield{author}{\bibinfo{person}{John~P. Smith~III},
  \bibinfo{person}{Andrea~A. diSessa}, {and} \bibinfo{person}{Jeremy
  Roschelle}.} \bibinfo{year}{1994}\natexlab{}.
\newblock \showarticletitle{{Misconceptions Reconceived: A Constructivist
  Analysis of Knowledge in Transition}}.
\newblock \bibinfo{journal}{\emph{The Journal of the Learning Sciences}}
  \bibinfo{volume}{3}, \bibinfo{number}{2} (\bibinfo{year}{1994}),
  \bibinfo{pages}{115--163}.
\newblock
\showISSN{1050-8406}
\href{https://doi.org/10.1207/s15327809jls0302{\_}1}{doi:\nolinkurl{10.1207/s15327809jls0302{\_}1}}


\bibitem[Sorva(2013)]%
        {Sorva2013NotionalEducation}
\bibfield{author}{\bibinfo{person}{Juha Sorva}.}
  \bibinfo{year}{2013}\natexlab{}.
\newblock \showarticletitle{{Notional machines and introductory programming
  education}}.
\newblock \bibinfo{journal}{\emph{ACM Transactions on Computing Education}}
  \bibinfo{volume}{13}, \bibinfo{number}{2} (\bibinfo{year}{2013}).
\newblock
\showISSN{19466226}
\href{https://doi.org/10.1145/2483710.2483713}{doi:\nolinkurl{10.1145/2483710.2483713}}


\bibitem[{Stack Overflow}(2025)]%
        {StackOverflow2025Survey}
\bibfield{author}{\bibinfo{person}{{Stack Overflow}}.}
  \bibinfo{year}{2025}\natexlab{}.
\newblock \bibinfo{title}{{2025 Stack Overflow Developer Survey}}.
\newblock \bibinfo{howpublished}{\url{https://survey.stackoverflow.co/2025/}}.
\newblock
\shownote{Accessed 2026-09-15}.
\newblock


\bibitem[Stray et~al\mbox{.}(2025)]%
        {Stray2025DeveloperStudy}
\bibfield{author}{\bibinfo{person}{Viktoria Stray},
  \bibinfo{person}{Elias~Goldmann Brandtz{\ae}g},
  \bibinfo{person}{Viggo~Tellefsen Wivestad}, \bibinfo{person}{Astri Barbala},
  {and} \bibinfo{person}{Nils~Brede Moe}.} \bibinfo{year}{2025}\natexlab{}.
\newblock \bibinfo{title}{{Developer Productivity With and Without GitHub
  Copilot: A Longitudinal Mixed-Methods Case Study}}.
\newblock \bibinfo{howpublished}{arXiv preprint arXiv:2509.20353}.
\newblock


\bibitem[Tang et~al\mbox{.}(2026)]%
        {Tang2026ProgrammingSessions}
\bibfield{author}{\bibinfo{person}{Ningzhi Tang}, \bibinfo{person}{Chaoran
  Chen}, \bibinfo{person}{Zihan Fang}, \bibinfo{person}{Gelei Xu},
  \bibinfo{person}{Maria Dhakal}, \bibinfo{person}{Yiyu Shi},
  \bibinfo{person}{Collin McMillan}, \bibinfo{person}{Yu Huang}, {and}
  \bibinfo{person}{Toby Jia-Jun Li}.} \bibinfo{year}{2026}\natexlab{}.
\newblock \bibinfo{title}{{Programming by Chat: A Large-Scale Behavioral
  Analysis of 11,579 Real-World AI-Assisted IDE Sessions}}.
\newblock \bibinfo{howpublished}{arXiv preprint arXiv:2604.00436}.
\newblock


\bibitem[Tankelevitch et~al\mbox{.}(2024)]%
        {Tankelevitch2024TheAI}
\bibfield{author}{\bibinfo{person}{Lev Tankelevitch}, \bibinfo{person}{Viktor
  Kewenig}, \bibinfo{person}{Auste Simkute}, \bibinfo{person}{Ava~Elizabeth
  Scott}, \bibinfo{person}{Advait Sarkar}, \bibinfo{person}{{Abigail Sellen}},
  {and} \bibinfo{person}{Sean Rintel}.} \bibinfo{year}{2024}\natexlab{}.
\newblock \showarticletitle{{The Metacognitive Demands and Opportunities of
  Generative AI}}. In \bibinfo{booktitle}{\emph{Proceedings of the 2024 CHI
  Conference on Human Factors in Computing Systems}}.
  \bibinfo{publisher}{Association for Computing Machinery}.
\newblock
\showISBNx{9798400703300}
\href{https://doi.org/10.1145/3613904.3642902}{doi:\nolinkurl{10.1145/3613904.3642902}}


\bibitem[Treude and Gerosa(2025)]%
        {Treude2025HowEngineering}
\bibfield{author}{\bibinfo{person}{Christoph Treude} {and}
  \bibinfo{person}{Marco~A. Gerosa}.} \bibinfo{year}{2025}\natexlab{}.
\newblock \bibinfo{title}{{How Developers Interact with AI: A Taxonomy of
  Human-AI Collaboration in Software Engineering}}.
\newblock \bibinfo{howpublished}{arXiv preprint arXiv:2501.08774}.
\newblock


\bibitem[Trippas et~al\mbox{.}(2024)]%
        {Trippas2024WhatWild}
\bibfield{author}{\bibinfo{person}{Johanne~R Trippas}, \bibinfo{person}{Sara
  Fahad~Dawood Al~Lawati}, \bibinfo{person}{Joel Mackenzie}, {and}
  \bibinfo{person}{Luke Gallagher}.} \bibinfo{year}{2024}\natexlab{}.
\newblock \showarticletitle{{What do Users Really Ask Large Language Models? An
  Initial Log Analysis of Google Bard Interactions in the Wild}}. In
  \bibinfo{booktitle}{\emph{Proceedings of the 47th International ACM SIGIR
  Conference on Research and Development in Information Retrieval}}
  \emph{(\bibinfo{series}{SIGIR '24})}. \bibinfo{publisher}{Association for
  Computing Machinery}, \bibinfo{address}{New York, NY, USA},
  \bibinfo{pages}{2703--2707}.
\newblock
\showISBNx{9798400704314}
\href{https://doi.org/10.1145/3626772.3657914}{doi:\nolinkurl{10.1145/3626772.3657914}}


\bibitem[Vaithilingam et~al\mbox{.}(2022)]%
        {Vaithilingam2022ExpectationModels}
\bibfield{author}{\bibinfo{person}{Priyan Vaithilingam},
  \bibinfo{person}{Tianyi Zhang}, {and} \bibinfo{person}{Elena~L. Glassman}.}
  \bibinfo{year}{2022}\natexlab{}.
\newblock \showarticletitle{{Expectation vs. Experience: Evaluating the
  Usability of Code Generation Tools Powered by Large Language Models}}. In
  \bibinfo{booktitle}{\emph{Extended Abstracts of the 2022 CHI Conference on
  Human Factors in Computing Systems (CHI EA '22)}}.
  \bibinfo{publisher}{Association for Computing Machinery}.
\newblock
\showISBNx{9781450391566}
\href{https://doi.org/10.1145/3491101.3519665}{doi:\nolinkurl{10.1145/3491101.3519665}}


\bibitem[Watanabe et~al\mbox{.}(2026)]%
        {Watanabe2026OnGitHub}
\bibfield{author}{\bibinfo{person}{Miku Watanabe}, \bibinfo{person}{Hao Li},
  \bibinfo{person}{Yutaro Kashiwa}, \bibinfo{person}{Brittany Reid},
  \bibinfo{person}{Hajimu Iida}, {and} \bibinfo{person}{Ahmed~E. Hassan}.}
  \bibinfo{year}{2026}\natexlab{}.
\newblock \showarticletitle{{On the Use of Agentic Coding: An Empirical Study
  of Pull Requests on GitHub}}.
\newblock \bibinfo{journal}{\emph{ACM Transactions on Software Engineering and
  Methodology}} \bibinfo{volume}{35}, \bibinfo{number}{10}
  (\bibinfo{year}{2026}), \bibinfo{pages}{1--24}.
\newblock
\href{https://doi.org/10.1145/3798166}{doi:\nolinkurl{10.1145/3798166}}


\bibitem[Weber et~al\mbox{.}(2024)]%
        {Weber2024SignificantModels}
\bibfield{author}{\bibinfo{person}{Thomas Weber}, \bibinfo{person}{Maximilian
  Brandmaier}, \bibinfo{person}{Albrecht Schmidt}, {and} \bibinfo{person}{Sven
  Mayer}.} \bibinfo{year}{2024}\natexlab{}.
\newblock \showarticletitle{{Significant Productivity Gains through Programming
  with Large Language Models}}.
\newblock \bibinfo{journal}{\emph{Proceedings of the ACM on Human-Computer
  Interaction}} \bibinfo{volume}{8}, \bibinfo{number}{EICS} (\bibinfo{date}{6}
  \bibinfo{year}{2024}).
\newblock
\showISSN{25730142}
\href{https://doi.org/10.1145/3661145}{doi:\nolinkurl{10.1145/3661145}}


\bibitem[Wen et~al\mbox{.}(2025)]%
        {Wen2025LANGUAGERLHF}
\bibfield{author}{\bibinfo{person}{Jiaxin Wen}, \bibinfo{person}{Ruiqi Zhong},
  \bibinfo{person}{Akbir Khan}, \bibinfo{person}{Ethan Perez},
  \bibinfo{person}{Jacob Steinhardt}, \bibinfo{person}{Minlie Huang},
  \bibinfo{person}{Samuel~R. Bowman}, \bibinfo{person}{He He}, {and}
  \bibinfo{person}{Shi Feng}.} \bibinfo{year}{2025}\natexlab{}.
\newblock \showarticletitle{{Language Models Learn to Mislead Humans via
  RLHF}}. In \bibinfo{booktitle}{\emph{The Thirteenth International Conference
  on Learning Representations (ICLR)}}.
\newblock
\urldef\tempurl%
\url{https://arxiv.org/abs/2409.12822}
\showURL{%
\tempurl}


\bibitem[Xiao et~al\mbox{.}(2024)]%
        {Xiao2024DevGPTConversations}
\bibfield{author}{\bibinfo{person}{Tao Xiao}, \bibinfo{person}{Christoph
  Treude}, \bibinfo{person}{Hideaki Hata}, {and} \bibinfo{person}{Kenichi
  Matsumoto}.} \bibinfo{year}{2024}\natexlab{}.
\newblock \showarticletitle{{DevGPT: Studying Developer-ChatGPT
  Conversations}}. In \bibinfo{booktitle}{\emph{Proceedings of the 21st
  International Conference on Mining Software Repositories}}
  \emph{(\bibinfo{series}{MSR '24})}. \bibinfo{publisher}{Association for
  Computing Machinery}, \bibinfo{pages}{227--230}.
\newblock
\href{https://doi.org/10.1145/3643991.3648400}{doi:\nolinkurl{10.1145/3643991.3648400}}


\bibitem[Zamfirescu-Pereira et~al\mbox{.}(2023)]%
        {Zamfirescu-Pereira2023WhyPrompts}
\bibfield{author}{\bibinfo{person}{J.~D. Zamfirescu-Pereira},
  \bibinfo{person}{Richmond~Y. Wong}, \bibinfo{person}{Bjoern Hartmann}, {and}
  \bibinfo{person}{Qian Yang}.} \bibinfo{year}{2023}\natexlab{}.
\newblock \showarticletitle{{Why Johnny Can't Prompt: How Non-AI Experts Try
  (and Fail) to Design LLM Prompts}}. In \bibinfo{booktitle}{\emph{Proceedings
  of the 2023 CHI Conference on Human Factors in Computing Systems (CHI '23)}}.
  \bibinfo{publisher}{Association for Computing Machinery}.
\newblock
\showISBNx{9781450394215}
\href{https://doi.org/10.1145/3544548.3581388}{doi:\nolinkurl{10.1145/3544548.3581388}}


\bibitem[Zhao et~al\mbox{.}(2024)]%
        {Zhao2024WildChat:Wild}
\bibfield{author}{\bibinfo{person}{Wenting Zhao}, \bibinfo{person}{Xiang Ren},
  \bibinfo{person}{Jack Hessel}, \bibinfo{person}{Claire Cardie},
  \bibinfo{person}{Yejin Choi}, {and} \bibinfo{person}{Yuntian Deng}.}
  \bibinfo{year}{2024}\natexlab{}.
\newblock \showarticletitle{{WildChat: 1M ChatGPT Interaction Logs in the
  Wild}}. In \bibinfo{booktitle}{\emph{The Twelfth International Conference on
  Learning Representations (ICLR)}}.
\newblock
\urldef\tempurl%
\url{https://arxiv.org/abs/2405.01470}
\showURL{%
\tempurl}


\bibitem[Zheng et~al\mbox{.}(2024)]%
        {Zheng2024LMSYSChat1M}
\bibfield{author}{\bibinfo{person}{Lianmin Zheng}, \bibinfo{person}{Wei-Lin
  Chiang}, \bibinfo{person}{Ying Sheng}, \bibinfo{person}{Tianle Li},
  \bibinfo{person}{Siyuan Zhuang}, \bibinfo{person}{Zhanghao Wu},
  \bibinfo{person}{Yonghao Zhuang}, \bibinfo{person}{Zhuohan Li},
  \bibinfo{person}{Zi Lin}, \bibinfo{person}{Eric~P. Xing},
  \bibinfo{person}{Joseph~E. Gonzalez}, \bibinfo{person}{Ion Stoica}, {and}
  \bibinfo{person}{Hao Zhang}.} \bibinfo{year}{2024}\natexlab{}.
\newblock \showarticletitle{{LMSYS-Chat-1M: A Large-Scale Real-World LLM
  Conversation Dataset}}. In \bibinfo{booktitle}{\emph{The Twelfth
  International Conference on Learning Representations (ICLR)}}.
\newblock
\urldef\tempurl%
\url{https://arxiv.org/abs/2309.11998}
\showURL{%
\tempurl}


\bibitem[Zi et~al\mbox{.}(2025)]%
        {Zi2025ICode}
\bibfield{author}{\bibinfo{person}{Yangtian Zi}, \bibinfo{person}{Luisa Li},
  \bibinfo{person}{Arjun Guha}, \bibinfo{person}{Carolyn Anderson}, {and}
  \bibinfo{person}{Molly~Q Feldman}.} \bibinfo{year}{2025}\natexlab{}.
\newblock \showarticletitle{{``I Would Have Written My Code Differently'':
  Beginners Struggle to Understand LLM-Generated Code}}. In
  \bibinfo{booktitle}{\emph{Companion Proceedings of the 33rd ACM International
  Conference on the Foundations of Software Engineering (FSE Companion '25)}}.
  \bibinfo{publisher}{Association for Computing Machinery},
  \bibinfo{pages}{1479--1488}.
\newblock
\showISBNx{9798400712760}
\href{https://doi.org/10.1145/3696630.3731663}{doi:\nolinkurl{10.1145/3696630.3731663}}


\bibitem[Ziegler et~al\mbox{.}(2022)]%
        {Ziegler2022ProductivityCompletion}
\bibfield{author}{\bibinfo{person}{Albert Ziegler}, \bibinfo{person}{Eirini
  Kalliamvakou}, \bibinfo{person}{X~Alice Li}, \bibinfo{person}{Andrew Rice},
  \bibinfo{person}{Devon Rifkin}, \bibinfo{person}{Shawn Simister},
  \bibinfo{person}{Ganesh Sittampalam}, {and} \bibinfo{person}{Edward
  Aftandilian}.} \bibinfo{year}{2022}\natexlab{}.
\newblock \showarticletitle{{Productivity assessment of neural code
  completion}}. In \bibinfo{booktitle}{\emph{Proceedings of the 6th ACM SIGPLAN
  International Symposium on Machine Programming}}.
  \bibinfo{publisher}{Association for Computing Machinery},
  \bibinfo{pages}{21--29}.
\newblock
\href{https://doi.org/10.1145/3520312.3534864}{doi:\nolinkurl{10.1145/3520312.3534864}}


\end{thebibliography}

\appendix
\section{The Preliminary Study}
\label{sec:preliminary-study}

This appendix documents the preliminary (JAWs) study \cite{OBrien2026UserAssistants} as provenance for the
present instrument: the misconception brainstorming exercise that generated
the candidate themes (Section~\ref{sec:brainstorming-provenance}), and the
coding procedure that turned them into the eleven-code preliminary codebook
and its validation round (Section~\ref{sec:preliminary-coding}). The
brainstormed themes are candidates, not findings: We make no claim that they
are prevalent, or that all of them occur.

\subsection{Brainstorming Provenance}
\label{sec:brainstorming-provenance}

The brainstorming began as a collaborative activity at the Copenhagen Symposium
on Human-Centered AI Adoption in Software Engineering (November 2024), an
international workshop supported by the Carlsberg Foundation and Alfred P.\ Sloan Foundation. Approximately 45 software engineering researchers from
institutions across North America, Europe, South America, and Oceania
participated, from mid-career to senior experts, representing diverse areas of expertise such as developer productivity, AI tool adoption,
and human factors in computing. The activity employed Liberating Structures
facilitation techniques to enable broad participation
\cite{McCandless2024LiberatingStructures}. Through structured small-group
discussions, participants shared observations from their research and teaching
experiences with programmers using LLM-based tools, identifying recurring
patterns of user confusion.

After the workshop, the author team worked asynchronously to systematize the
brainstormed misconceptions, drawing on both the workshop discussions and
peer-reviewed literature, and incorporating new literature as we encountered
it. We organized the material into themes through an iterative process of
grouping similar concepts and refining category definitions. Claims derive from
three distinguishable sources, marked in Table~\ref{tab:brainstormed-themes}:
peer-reviewed studies (cited), observations contributed during the workshop
discussion, and experiences reported by individual authors. The latter two are
anecdotal by construction. We include them because they offer candidates the published literature has not yet cataloged.

Through our discussions, we observed that some misconceptions concern
properties of large language models as a class---holding for any system built
on one---whereas others concern properties of a particular deployed tool, such
as whether it can search the web or recall a previous session. The latter are
configuration choices that vary across providers, versions, and subscription
tiers, so whether a given belief is mistaken depends on the deployment.
Table~\ref{tab:brainstormed-themes} presents the themes following this
distinction.

This exercise is in no way exhaustive, and we do not intend it to be: The set of potential misconceptions is infinitely large, but not all are necessarily
interesting. We focused, as in the definition of programmers' misconceptions,
on beliefs that we would expect to impact the choices a programmer makes about
how to use an LLM-based assistant \cite{Sorva2013NotionalEducation}. The themes
below seeded the initial codebook applied in the preliminary study. As described in Section~\ref{sec:research-approach}, only categories with clear,
observable traces in conversation logs survived into the final eight-code
instrument.

\begin{table}[t]
\caption{Misconception themes generated in the preliminary brainstorming
exercise. Sources: peer-reviewed literature (cited); \textsuperscript{W}
observation contributed in workshop discussion; \textsuperscript{A} experience
reported by an individual author. Themes marked \textsuperscript{W} or
\textsuperscript{A} without citation are anecdotal.}
\label{tab:brainstormed-themes}
\footnotesize
\begin{tabular}{p{2.6cm}p{7.6cm}p{2.6cm}}
\toprule
\textbf{Theme} & \textbf{Candidate misconception} & \textbf{Sources} \\
\midrule
\multicolumn{3}{l}{\textit{About a particular deployed tool}} \\
\midrule
Information retrieval mechanism & The tool retrieves answers from a database
or by keyword lookup; confusion about whether a tool consults
retrieval-augmented sources before generating, and which knowledge bases those
include. &
\cite{OBrien2025HowProgram, Nguyen2024HowOther, Feldman2024Non-ExpertFuture,
Brachman2025BuildingChat-bot}\textsuperscript{A} \\
Agentic actions & The tool takes agentic actions by default---searching the
web, executing programs, or proactively validating code---when it lacks these
affordances. &
\cite{Prather2023ItsProgrammers, OBrien2025HowProgram} \\
Session Memory & Information from previous conversations is available in the
current session; code versions within a session are checkpointed and can be
restored by request. &
\cite{Zamfirescu-Pereira2023WhyPrompts} \\
Session persistence & The tool continues ``processing'' a task after the
conversation ends, such that a user can return later for results. &
\textsuperscript{A} \\
Scope of access & Uncertainty about which local files and directories an
environment-integrated tool can read, or about how to hide information from
it (e.g., that inline comments are ignored). &
\textsuperscript{W} \\
Continuous Training & The tool's knowledge has no training cutoff, or the
tool learns from the user's habits and feedback during use. &
\cite{Zamfirescu-Pereira2023WhyPrompts} \\
Deterministic behavior & Identical prompts always yield identical responses. &
\cite{Lau2023FromCopilot, Nguyen2024HowOther} \\
Model family & A conversational interface is assumed to be LLM-based when it
is not (or vice versa), since rule-based systems can present visually
indistinguishable interfaces. &
\cite{Brachman2025BuildingChat-bot, BrachmanFollowSystems,
Do2024EvaluatingSystems} \\
\midrule
\multicolumn{3}{l}{\textit{About language models as a class}} \\
\midrule
Stability of results & Outputs are not expected to change under semantically
irrelevant prompt modifications, such as added whitespace or distractor
sentences. &
\cite{Mirzadeh2024GSM-Symbolic:Models, Zamfirescu-Pereira2023WhyPrompts,
Jayagopal2022ExploringProgrammers} \\
Groundedness & Some circumstances make hallucination impossible---for
example, about an uploaded data file, or about meta-information such as
``Which model are you?'' &
\textsuperscript{A} \\
Native explainability & A model's explanation of code it generated faithfully
reports the reasoning that produced the code. &
\cite{Lehtinen2024LetsQuestions, Ferdowsifard2020Small-stepExample}\textsuperscript{A} \\
Symbolic logic & Correct answers to algebra or logic problems are produced by
a calculator-like mechanism, rather than by generation with unintuitive
sensitivities to problem presentation. &
\cite{Mirzadeh2024GSM-Symbolic:Models} \\
Context window & Providing more information always improves performance
(``monotonicity belief''); unawareness of context limits and of position
effects on recall. &
\cite{Jayagopal2022ExploringProgrammers, Tankelevitch2024TheAI,
Liu2024LostContexts} \\
\bottomrule
\end{tabular}
\end{table}

\subsection{Preliminary Coding Procedure}
\label{sec:preliminary-coding}

Starting from the brainstormed themes, we drafted an initial codebook
applicable to real conversations. Not all themes were relevant to the WildChat assistant. Some concern only specific architectures, such as retrieval-augmented
generation. Two authors (GO and APSA) applied the initial codebook, each coding 250 unique conversations from the 500-conversation random sample, with multiple non-exclusive labels allowed per conversation. During this coding they added emergent codes for notable observations, plus bookkeeping codes
for conversations that were mostly non-English (85, likely automated
language-detection errors), lacked discernible user instructions, or were not
Python-related (7, reflecting limitations of the regular-expression filter);
these were not coded for misconceptions.

Because potential misconceptions were a minority class, the two coders then reviewed each other's flagged conversations (78 in total) and discussed
disagreements, rather than reviewing the full sample jointly. We performed
axial coding to finalize the codes---de-duplicating, merging insufficiently
distinct codes, and removing codes from the initial codebook that had not been
observed---and a third author (MK) adjudicated outstanding disagreements. The
result was the eleven-code preliminary codebook shown in
Table~\ref{tab:preliminary-codebook}, with adjudicated labels on the 50
conversations where at least two coders agreed that a potential misconception
was present.

We then conducted a validation round with three authors of the workshop version
not involved in the initial coding (SB, GL, and Mircea Lungu). We drew a stratified subset of
\MetPrelimConvs{} conversations from the flagged set, ensuring at least one
example of each code. Validation coders could apply multiple codes per
conversation and knew that each conversation had been flagged as a potential
misconception by at least two members of the coding team. One validation coder
revised their labels after a group discussion. We use the revised set.
Agreement across the four resulting label sets---the coding team's adjudicated
labels and the three validation coders---is reported in Section~\ref{sec:research-approach}.

Reviewing annotation notes after the validation round, we found the
disagreement causes described in Section~\ref{sec:research-approach} (i.e., difficult-to-parse language, pasted coursework, and long transcripts). One
further pattern was a difference in labeling thresholds: The coding team had
flagged any conversation where a misconception seemed plausible, while
validation coders tended to reserve labels for clear-cut cases---even though
all coders endorsed the codebook itself as reasonable. This threshold
difference, as much as any disagreement about definitions, motivated the
boundary-case rules built into the revised codebook
(Section~\ref{sec:research-approach}).

Of the eleven preliminary codes, eight were carried into the revised
instrument (Web Access continuing as Internet Access, and Clear Chat as Clear
Session). Dynamic Analysis, Restrict Sources, and Other Agentic Behavior were dropped on the observability grounds described in Section~\ref{sec:research-approach}.

\begin{table}[t]
\caption{The first version of the codebook, as used in the preliminary
study. Codes are non-exclusive. Per-code frequencies are reported in the
workshop version of this study \cite{OBrien2026UserAssistants} and are not
repeated here.}
\label{tab:preliminary-codebook}
\footnotesize
\begin{tabular}{p{2.5cm}p{6.0cm}p{4.3cm}}
\toprule
\textbf{Code} & \textbf{Definition} & \textbf{Example} \\
\midrule
Web Access & The prompt asks the LLM to access information located at a
particular URL, such as data or code, or requires knowledge of the HTML of a
website in the context of web scraping. & ``Write me a webscraper that
will pull the top 100 most popular wikipedia page titles and their visit
counts and save it in a csv''. \\
Dynamic Analysis & The prompt asks for specific code-checking actions that
cannot be fulfilled via static text analysis alone (e.g., runtime errors in
Python). & ``Remove all errors from this program.'' \\
Algebra & The prompt asks for an algebraic statement to be evaluated, or
would require one to fulfill this. & ``What is the value of $\langle$algebraic
expression$\rangle$?'' \\
Code Execution & The prompt requests the output of running a program or asks
for the program to be executed. & ``Run this program and tell me the
output.'' \\
Session Memory & The prompt refers to information from a separate
conversation as if the LLM has access to that conversation. & ``Can you
add $\langle$new feature$\rangle$ to the code you suggested in our last
conversation?'' \\
Non-Text Output & The prompt asks the LLM to return something that is not
text, such as an image. & ``Give me a graph of $\langle$company$\rangle$
stock value for the last X years.'' \\
Local Machine Access & The prompt asks for other software tools to be run on
the user's machine. & ``Can you run this in my terminal?'' \\
Restrict Sources & The prompt asks the model to limit the kinds of
information or sources used to generate the output. & ``Only use code
from $\langle$repository address$\rangle$ to come up with your answer.'' \\
Continuous Training & The prompt asks the LLM to use the ``latest'' or ``up
to date'' information. & ``Write a script using the latest version of
$\langle$library$\rangle$.'' \\
Clear Chat & The prompt asks the model to delete information or clear the
chat. & ``Clear the chat.'' \\
Other Agentic Behavior & The prompt asks the LLM to take an agentic action,
such as asking the model to take a screenshot, play a sound, or use software
on an unspecified machine. & ``Using visual studio code ...'' \\
\bottomrule
\end{tabular}
\end{table}

\section{Code Boundary Conditions}
\label{sec:code-boundaries}

Table~\ref{tab:code-boundaries} reproduces the boundary conditions from the
final codebook: for each code, the cases annotators were instructed
\emph{not} to count as the misconception.

\begin{table*}[t]
\caption{Boundary conditions from the final codebook: for each code, the cases annotators were instructed NOT to count as the misconception. Rules under ``All codes'' apply to every code.}
\label{tab:code-boundaries}
\small
\begin{tabular}{p{0.16\textwidth}p{0.76\textwidth}}
\toprule
Code & Excluded boundary cases \\
\midrule
All codes & Prompts that appear to be copy-pasted jailbreak or persona templates: capabilities asserted by the template are not treated as the user's own beliefs. \\
\midrule
Algebra & Asking for Python code that evaluates the expression --- deliverable is code. \par Explaining an algebraic identity or deriving a formula conceptually. \par Implementing a mathematical algorithm in code. \par Whole-conversation rule: if the user appears satisfied with a code response, default to No Misconception. This includes a user who first asks for the answer to a math task but then asks for code for the task and accepts it. \par Asking the LLM to write or provide a formula/equation (rather than evaluate or solve it) is not Algebra --- e.g. asking the model to write the equation that would solve for a quantity, without computing the value. \\
\midrule
Clear Session & Conversational turns initiating a new direction for the chat: ``Ignore what I just said'', ``let's start over with a new approach'' \\
\midrule
Code Execution & Asking the LLM to trace code by hand or predict output (``walk through this loop for n=3'') \par Asking for code that implements a computation. \par Whole-conversation rule: if the user appears satisfied with a code response, default to No Misconception. \\
\midrule
Continuous Training & Specifying a known stable version (``use numpy 1.20'') is not a misconception. \\
\midrule
Internet Access & Vague phrases like ``modern UI off the internet'' --- default to No Misconception. \par A URL appearing only inside pasted code, tracebacks, or error messages. \par Requests for a generic scraper template against an unspecified site. \par Placeholder or template URLs (e.g., example.com, or a URL inside a format/JSON example the user supplies) --- the task does not depend on fetching them. \par A link appearing incidentally in a longer task description (e.g., a reference page about a concept or topic mentioned in an assignment) that the prompt's instructions never invoke. \\
\midrule
Local Machine Access & Asking for instructions that can be performed by the user on their  local machine: ``How do I find what version of Python I have installed?''  ``How do I install X?'' \par Pasting a local file path as context for debugging without asking the LLM to access it (e.g., ``the file is here: C:\textbackslash{}Users\textbackslash{}...'' within a debugging conversation). \\
\midrule
Non-Text Output & ``Write Python code that produces a plot of X'' --- deliverable is code, not the plot. \par Whole-conversation rule: if the user appears satisfied with a code response, default to No Misconception. This includes accepting the code implicitly, e.g. by continuing to refine the request in follow-up turns without objecting to having received code. \\
\midrule
Session Memory & Referencing prior conversations, but including relevant artifacts from those conversations as if the assistant cannot recall on its own. For example, ``Remember what you wrote yesterday...'' followed by a code block, or citing a previous answer while pasting that answer in full in the same prompt. \par References to artifacts never shared in the conversation WITHOUT an explicit cross-session anchor (``last time'', ``previous chat'', ``yesterday'') --- could be a user mistake or a forgotten paste; default to No Misconception. Flag only when an explicit cross-session anchor is present AND the user does not re-post the answer or context their prompt requires. \\
\bottomrule
\end{tabular}
\end{table*}

\section{Pipeline Details}
\label{sec:pipeline-details}

This appendix reports the operational detail needed to assess and reproduce
the model-assisted pipeline stages (Section~\ref{sec:methods}), following
published reporting guidelines for empirical software engineering studies
involving large language models \cite{Baltes2025GuidelinesModels}. The complete
prompts and run scripts for both LLM stages are included in the replication
package (\url{https://github.com/elleobrien/wildchat-misconceptions-replication})
rather than reproduced here.

\subsection{Models and Endpoints}

We used three models, each in a distinct role. \verb|gpt-oss-120b|, an
open-weight model, performed the full-corpus screening pass. It ran under Ollama on university high-performance computing GPUs, serving the MXFP4 weights, with reasoning effort left at the model default.
\verb|claude-sonnet-4-6| re-scored the screening candidates, and
\verb|claude-opus-4-7| applied the finalized codebook in the annotation stage. We accessed both through a university-hosted, OpenAI-compatible API gateway (the U-M GPT Toolkit); extended thinking was not enabled. This division of labor is
deliberate: The stage that touches every conversation in the corpus --- and
therefore has the largest influence on what enters the study --- uses an
open-weight model, so it can be reproduced without depending on a hosted
service that may change or be withdrawn.

\subsection{Prompt Structure}

Both LLM stages share the same call structure: Each conversation is scored once per code, so a conversation--code cell corresponds to one API call. The
user message presents the code's criteria, then the full transcript wrapped
in \verb|<conversation>| tags, followed by an instruction that the wrapped
text is data to analyze rather than instructions to follow (a guard against
prompt injection from transcript contents). We require the model to produce its reasoning \emph{before} its verdict, ending with a final line containing
only the digit 0 or 1.

The two stages differ in their system prompts and per-code prompt text,
which encode opposite error preferences. The screening (filter) track
instructs the model to err on the side of labeling: False positives are declared acceptable and missed positives are not, and any reasonable reading
under which the criteria fit should be labeled 1. The annotation track
instead instructs the model that it is scoring candidates from a
deliberately over-permissive filter, to apply the codebook definition
``accurately, not generously,'' to apply boundary cases and exclusions
strictly, and to read the whole conversation before deciding (implementing
the codebook's whole-conversation rules). The per-code prompts in the
annotation track are built mechanically from the human-facing codebook columns (i.e., definition, examples, and boundary cases), so the instrument the
model applies is the same one the human annotators used. Both system prompts and all sixteen per-code prompts (eight per track) are included verbatim in the replication package.

\subsection{Sampling Parameters and Run Stability}

No sampling parameter (e.g., temperature, top-$p$, or token limit) is set anywhere in the pipeline code. Every call used the default parameters of its serving endpoint. Per the endpoints' documentation (retrieved 2026-08-25), the
applicable default was temperature 1.0 at every stage: The Ollama distribution of \verb|gpt-oss-120b| packages temperature 1.0 in its
Modelfile, and the Anthropic Messages API defaults to temperature 1.0 when none is supplied. Consistent with non-zero temperature, the pipeline is not perfectly deterministic: In a repeated-run check of the screening prompts on the anchor set (run through the gateway's Bedrock-hosted copy of \verb|gpt-oss-120b| rather than the HPC Ollama deployment used for the corpus screen), all \MetAnchorConvs{} anchor conversations were recovered in
each of 10 independent runs, but the total number of flagged
conversation--code cells varied between 17 and 20 of 96 (mean 18.3).

\subsection{Inference Dates, Calls, and Token Totals}

The screening pass comprised \MetScreenTotalCalls{} calls
(\MetCorpusConvs{} conversations $\times$ 8 codes) and ran on university HPC
resources in June 2026, with no per-token cost. The confirmation pass ran on
June 11--15, 2026, and consumed \MetConfTokIn{} input and \MetConfTokOut{}
output tokens across \MetConfCalls{} calls. The annotation pass completed on
July 30, 2026, and consumed \MetAnnoTokIn{} input and \MetAnnoTokOut{}
output tokens across \MetAnnoCalls{} calls. The Internet Access re-run followed on August 4, 2026. At list prices at the time, the Sonnet re-score and the Opus annotation pass each cost roughly \$40.

\subsection{Screening Model Selection}

As described in Section~\ref{sec:methods}, we developed the filter prompts against \verb|claude-opus-4-7| and screened the corpus with a less expensive model. We evaluated five candidate screening models on
the \MetAnchorConvs{}-conversation anchor set: \verb|gpt-oss-120b|, Gemma
3 27B, GPT-4o-mini, and Llama 4 Scout and Maverick. Two other candidates also recovered every anchor, but only by flagging most of the anchor-set cells (Gemma 3 27B, 82 of 96; GPT-4o-mini, 55 of 96), and the Llama 4 models missed one or two anchors. We selected \verb|gpt-oss-120b| as the only candidate with full recall, a selective flag rate, and open weights. The choice of annotation model is reported in Section~\ref{sec:methods}.

\subsection{Label Verification Tool}
\label{sec:verification-tool}

\begin{figure}[htbp]
  \centering
  \includegraphics[width=\linewidth]{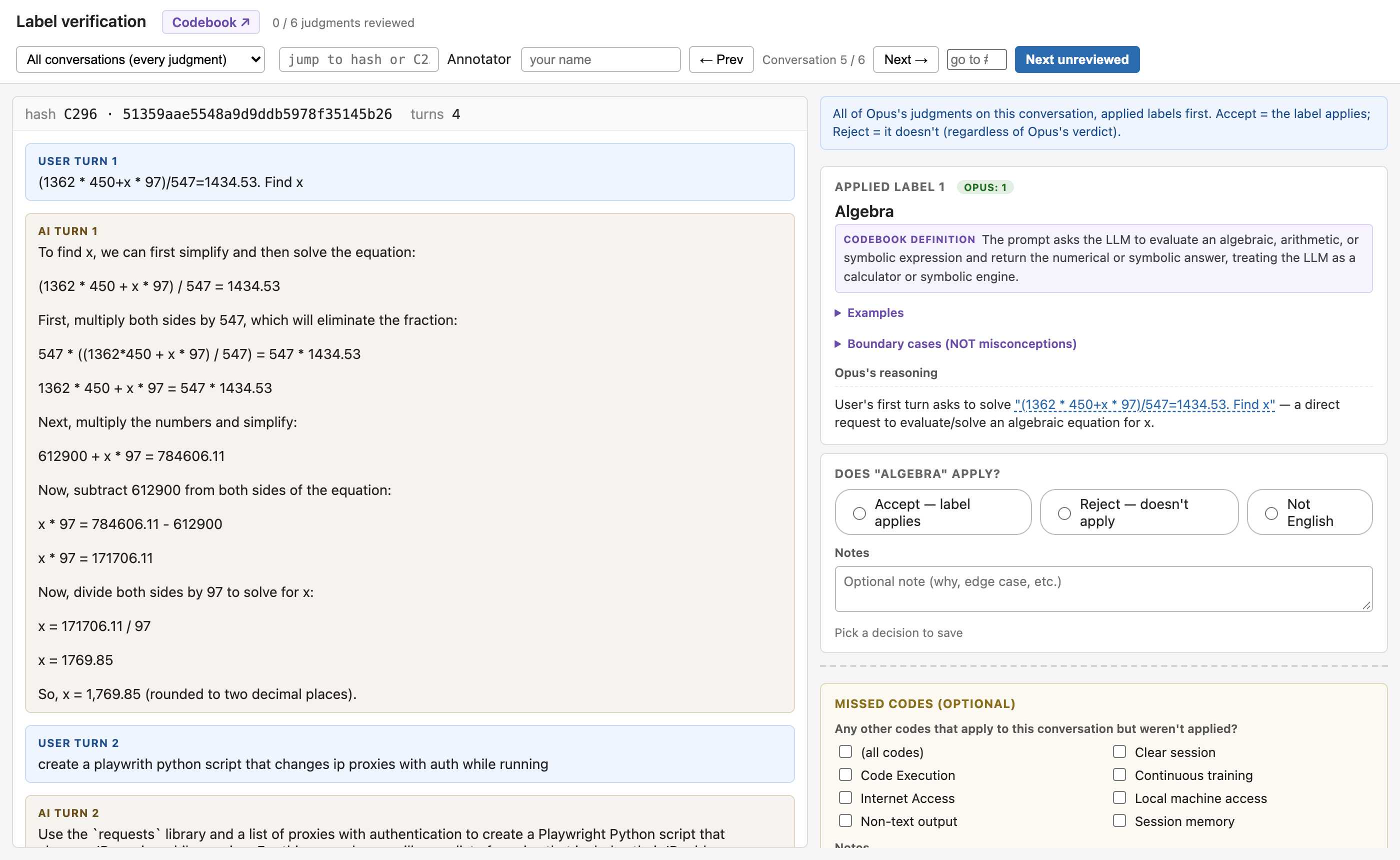}
  \Description{Screenshot of a web application. The left panel shows a conversation transcript in alternating user and AI turns; the first user turn asks the model to solve an equation for x. The right panel shows a card for the model-applied Algebra label, with the codebook definition, collapsible examples and boundary cases, and the model's reasoning quoting the user's prompt. Below it are buttons to accept or reject the label or mark the conversation as not English, a notes field, and a checklist of other codes the annotator can add.}
  \caption{The label verification tool, showing conversation C296 (Section~\ref{sec:results}), to which the annotation model applied the Algebra code.}
  \label{fig:annotation-tool}
\end{figure}

Figure~\ref{fig:annotation-tool} shows the browser-based tool used for human
verification of the model-applied labels (Section~\ref{sec:methods}). The
left panel shows the full transcript. The right panel lists every judgment
the annotation model made on the conversation, each with the codebook
definition (with collapsible examples and boundary cases) and the model's
reasoning, in which quoted passages from the transcript are underlined. The
annotator accepts or rejects each judgment, can mark the conversation as not
English, can leave a note, and can apply any code the model missed.

\section{Supplementary figures}\label{sec:supplementary-figures}

We regenerated the figures in this appendix from the final verified label set (2026-08-25). Scripts and data are included in the replication package.

\begin{figure}[htbp]
  \centering
  \includegraphics[width=0.85\linewidth]{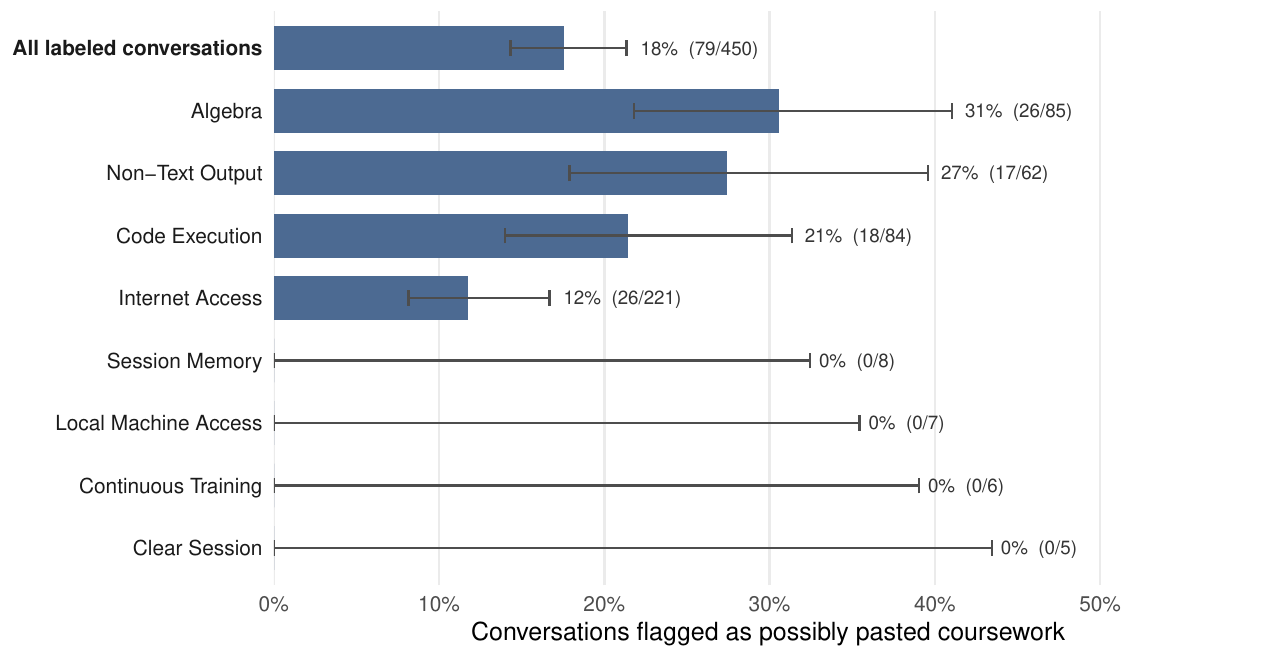}
  \Description{Horizontal bar chart with one bar for all labeled conversations and one bar per misconception code, showing the share of conversations flagged as possibly pasted coursework, with 95\% confidence intervals. Algebra and Non-Text Output have the highest shares, Code Execution is close to the overall share, and Internet Access is lower. The four rarest codes have no flagged conversations and wide intervals.}
  \caption{Share of labeled conversations flagged as possibly pasted coursework, by misconception code. The flag is the union of the lexical coursework measure (Section~\ref{sec:results}; exact patterns in the replication package) and annotator coursework marks recorded during label verification. Error bars are 95\% Wilson intervals; the omnibus chi-squared across codes (simulated $p$, due to small cells) gives $\chi^2$ = \ResHWChiSq{}, $p$ = \ResHWP{}, Cram\'er's $V$ = \ResHWV{} (\ResHWRows{} conversation--label rows). Coursework traces concentrate in the Algebra and Non-text output categories.}
  \label{fig:hw-by-code}
\end{figure}

\begin{figure}[htbp]
  \centering
  \includegraphics[width=\linewidth]{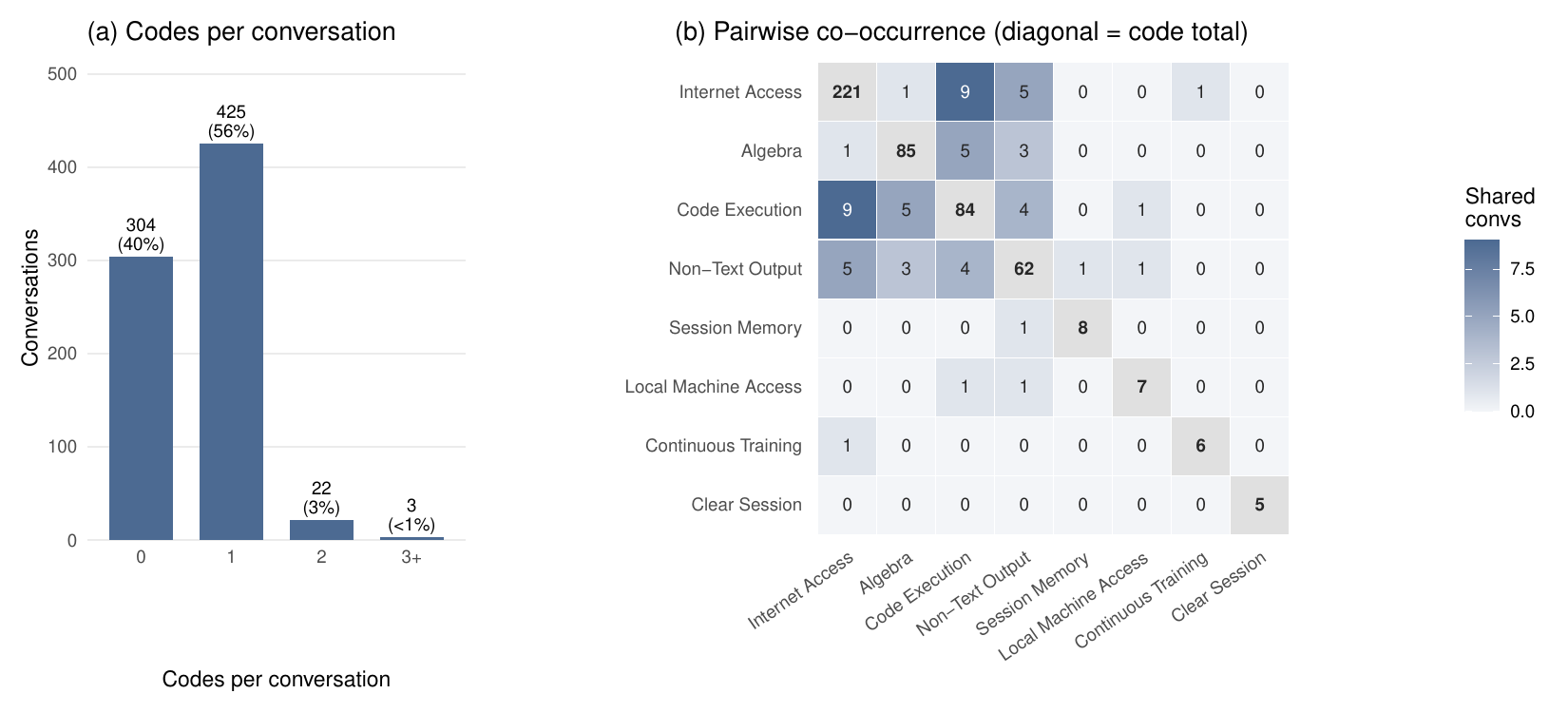}
  \Description{Two panels. Panel (a) is a bar chart of the number of misconception codes per reviewed conversation: most labeled conversations have exactly one code, and few have two or more. Panel (b) is a heatmap of pairwise code co-occurrence, with code totals on the diagonal. Off-diagonal counts are small; the largest is between Internet Access and Code Execution.}
  \caption{Label multiplicity in the analysis corpus. (a) Number of misconception codes per reviewed conversation; most labeled conversations carry exactly one code. (b) Pairwise co-occurrence of codes (diagonal = code totals). The largest overlap is between Internet Access and Code Execution (9 conversations), typically requests to both fetch a resource and run code against it.}
  \label{fig:code-cooccurrence}
\end{figure}

\section{Sensitivity analyses}\label{sec:sensitivity}

The omnibus tests in Sections~\ref{sec:results} and~\ref{sec:machine-responses} are chi-squared tests that treat each conversation, or each IP--code pair, as an independent observation. Three features of the data could violate that assumption. First, the same hashed IP can contribute several conversations (\SensNIPMulti{} of \SensNIP{} IPs did, contributing \SensNConvFromMulti{} of the \SensNConv{} labeled conversations). Second, a conversation carrying more than one code contributes one row per code to the refusal-by-label table. Third, an IP with many conversations has more opportunities to exhibit a code twice, so recurrence-by-code could reflect user activity rather than persistence of the belief. Table~\ref{tab:sensitivity} repeats each test with an adjustment for the corresponding feature. All of these tests are exploratory and are reported without correction for multiple comparisons; effect sizes accompany each in the main text. The script (\texttt{sensitivity\_clustering.R}) is in the replication package.

\begin{table}[htbp]
  \centering
  \caption{Main chi-squared tests and their adjusted counterparts. Mixed-effects models are logistic regressions fitted with \texttt{lme4}; each adjusted $p$ is a likelihood-ratio test of the model with the predictor against the same model without it.}
  \label{tab:sensitivity}
  \small
  \resizebox{\linewidth}{!}{\begin{tabular}{p{0.17\linewidth} p{0.24\linewidth} r p{0.38\linewidth} r}
\toprule
Test & Unadjusted & $p$ & Adjusted & $p$ \\
\midrule
Refusal by model family & $\chi^2$, 450 conversations & 0.049 & Mixed-effects logistic, IP random intercept (LRT) & 0.079 \\
Refusal by misconception label & $\chi^2$ (simulated), 478 conversation--label rows & $1\times10^{-5}$ & Mixed-effects logistic, IP random intercept (LRT) & $2\times10^{-6}$ \\
 & & & Single-label conversations only (425), IP random intercept (LRT) & $3\times10^{-5}$ \\
Recurrence by code & $\chi^2$ (simulated), 353 IP--code pairs & 0.002 & Logistic regression with log(conversations per IP) covariate (LRT) & 0.001 \\
\bottomrule
\end{tabular}
}
\end{table}

\paragraph{IP clustering.} We refit the two refusal-rate comparisons as mixed-effects logistic regressions with a random intercept per hashed IP. The model-family difference reported in Section~\ref{sec:machine-responses} did not survive this adjustment (odds ratio for GPT-4 versus GPT-3.5 \SensFamOR{}, 95\% Wald CI [\SensFamORLo{}, \SensFamORHi{}], $p$ = \SensFamGlmerP{}), and we do not interpret it. The difference across labels was unchanged ($\chi^2$ = \SensLabGlmerChiSq{} on \SensLabGlmerDf{} $df$, $p$ = \SensLabGlmerP{}).

\paragraph{Label multiplicity.} Restricting the refusal-by-label test to conversations carrying exactly one code (\SensSingleN{} of \SensNConv{}), with the IP random intercept, gave $\chi^2$ = \SensSingleGlmerChiSq{}, $p$ = \SensSingleGlmerP{}.

\paragraph{User activity.} For each IP--code pair we recorded the IP's total number of conversations in the full English--Python corpus (\MetCorpusConvs{} conversations, before candidate flagging). Activity differed across codes (IPs asking for non-text output were the least active) and predicted recurrence, so we fit a logistic regression of recurrence on code with log activity as a covariate. We included a random intercept for IP in a first fit, but its variance estimate was zero, so we report the ordinary logistic regression. Adding code to the activity-only model improved fit ($\chi^2$ = \SensRecAdjChiSq{} on \SensRecAdjDf{} $df$, $p$ = \SensRecAdjP{}).

\end{document}